\documentclass[aps,prd,amsmath,floats,floatfix,twocolumn,superscriptaddress,nofootinbib]{revtex4}
\usepackage{graphicx,epsfig,amssymb,amsmath,amsbsy,mathrsfs,widetext}
\usepackage{bm,array}

\usepackage[usenames]{color}
\usepackage{ulem}
\allowdisplaybreaks
\newcommand{\beq}{\begin{equation}}
\newcommand{\eeq}{\end{equation}}
\newcommand{\beqa}{\begin{eqnarray}}
\newcommand{\eeqa}{\end{eqnarray}}

\newcommand{\SMetric}{g}     % spatial metric
\newcommand{\SRicci}{R}      % spatial Ricci tensor
\newcommand{\SRicciS}{R}     % spatial Ricci scalar
\newcommand{\ExCurv}{K}      % spatial extrinsic curvature
\newcommand{\TrExCurv}{K}    % trace of spatial extrinsic curvature
\newcommand{\CF}{\psi}              % conformal factor

\newcommand{\CMetric}{{\tilde{g}}}    % conformal spatial metric
\newcommand{\CITA}{\affiliation{Canadian Institute for Theoretical Astrophysics,
60 St. George Street, University of Toronto, Toronto, ON M5S 3H8, Canada}}

\begin{document}
\vspace{-2.5cm} 

\title{Including realistic tidal deformations in binary black-hole initial data}

\author{Tony Chu} \CITA

\date{\today}

\begin{abstract}
A shortcoming of current binary black-hole initial data
is the generation of spurious gravitational radiation, 
so-called junk radiation, when they are evolved. 
This problem is a consequence of an oversimplified modeling of the
binary's physics in the initial data.
Since junk radiation is not astrophysically realistic, it contaminates 
the actual waveforms of interest and poses a numerical nuisance.
The work here presents a further step towards mitigating and understanding 
the origin of this issue, by
incorporating post-Newtonian results in the construction of
constraint-satisfying binary black-hole initial data.
Here we focus on including realistic tidal deformations of 
the black holes in the initial data, by building on the method of 
superposing suitably chosen black hole metrics to compute the conformal data.
We describe the details of our initial data for an equal-mass and 
nonspinning binary, compute the subsequent relaxation 
of horizon quantities in evolutions, and quantify the amount of junk 
radiation that is generated. These results are contrasted with those 
obtained with the most common choice of conformally flat (CF) initial data, 
as well as superposed Kerr-Schild (SKS) initial data. 
We find that when realistic tidal deformations are included, 
the early transients in the horizon geometries are significantly reduced, 
along with  smaller deviations in the relaxed black hole masses and spins 
from their starting values. Likewise, the junk radiation content 
in the $l=2$ modes is reduced by a factor of $\sim$1.7 relative to 
CF initial data, but only by a factor of $\sim$1.2 relative to SKS 
initial data. More prominently, the junk radiation content in the 
$3\leq l\leq8$ modes is reduced by a factor of $\sim$5 relative to CF 
initial data, and by a factor of $\sim$2.4 relative to SKS initial data. 
\end{abstract}

\maketitle

%#########################
\section{Introduction}
%#########################
A key objective of numerical relativity is to accurately model the
inspiral and coalescence of black-hole binaries, which are important
sources of gravitational waves that are expected to be observed by
detectors such as LIGO~\cite{ligoBBH} and VIRGO~\cite{Acernese:2002}
in the near future. Any simulation of a black-hole binary must begin
with the construction of suitable initial data, which are a solution to the
Einstein constraint equations, that ideally capture as many relevant
features of the physical system as possible. Presently though, the
majority of initial data assumes that the spatial metric is conformally
flat, a choice dictated by convenience. It is known that conformal
flatness is generally incompatible with desired black hole solutions.
For instance, a Kerr black hole with nonzero spin does not admit a
conformally flat slicing~\cite{GaratPrice:2000,Kroon:2004b}, and neither does
a black-hole binary starting at ${\mathcal O\left(v^4\right)}$
in the post-Newtonian (PN) approximation~\cite{Nissanke2006},
where $v$ is the binary's orbital velocity, in units of the speed of 
light $c$.

One side effect of conformally flat initial data for black-hole binaries
is the generation of
spurious gravitational radiation, so-called junk radiation, when they
are evolved.
Junk radiation contaminates the waveforms of interest, 
and interferes with their comparison to PN
predictions~\cite{Boyle2007,Hannam2007}. Computational resources
are also wasted in waiting for the junk radiation to propagate
off the computational domain, before reliable waveforms can be extracted.
Resolving the high-frequency components of the junk radiation 
requires a large increase in numerical resolution as 
well~\cite{Zlochower2012}, which slows 
down the evolution appreciably.
The initial properties of the black holes themselves are also
altered by the junk radiation, relaxing to slightly different values
later on in the evolution.
Additionally, junk radiation disturbs the orbital trajectories and 
complicates the construction of initial data with low eccentricity, 
which applies to binaries formed from stellar 
evolution~\cite{Postnov:2007jv}. 
Recently though, an iterative scheme was demonstrated to be effective at 
jointly reducing eccentricity and junk radiation~\cite{Zhang:2013gda}.

Over the last several years, various efforts have been made to go beyond
the assumption of conformal flatness, by using conformally curved initial
data. A direct superposition of black hole metrics was introduced
in~\cite{Matzner1999,Marronetti-Matzner:2000,Marronetti2000} to specify
the conformal metric, and a similar procedure was shown in~\cite{Hannam2007b}
to reduce the junk radiation in the head-on collision of two black holes.
Later on, a weighted superposition of black hole metrics was used
in~\cite{Lovelace2008,Lovelace2009}, and decreased the amount of junk
radiation in the inspiral of two equal-mass, nonspinning black
holes~\cite{Lovelace2009}. These superposed black-hole initial data already
provide a notable improvement over conformally flat initial data, but
they do not take advantage of all the available information to
better represent the binary's physics, such as results from PN
theory~\cite{Blanchet2006}. Including such information could prove to
be very useful.

Initial data incorporating the PN approximation include that
of~\cite{Nissanke2006}, which has interaction terms between
the black holes in the conformal metric, and that of~\cite{kellyEtAl:2007},
which contains the outgoing gravitational radiation of the binary
in the conformal metric. 
The initial data in~\cite{kellyEtAl:2007} was evolved
in~\cite{Kelly2010,Mundim2011}, and was found to reduce the
low-frequency components of the junk radiation. However, currently such 
initial data are largely restricted to nonspinning black holes.
Furthermore, the regions near the black holes are not adequately treated in
the approaches above, because no attempt was made to account for the
tidal deformations.

The aforementioned issue was addressed in~\cite{JohnsonMcDaniel:2009dq},
by describing the vicinity of the black holes by tidally perturbed
Schwarzschild metrics in horizon penetrating coordinates,
which were asymptotically matched to a PN metric to determine the tidal fields. 
This procedure has now been extended to spinning black 
holes~\cite{Gallouin:2012}.
These studies follow the earlier work of~\cite{Yunes2006a,Yunes2006b},
which used black hole metrics in coordinates that were not horizon
penetrating from the outset, and were thus inconvenient for numerical 
implementation.
Including tidal deformations is expected to reduce the 
high-frequency components of the junk radiation, which are 
typically attributed to physically unrealistic deformations of the black holes 
in the initial data that radiate away as the black hole geometries 
relax in an evolution.
The initial data of~\cite{JohnsonMcDaniel:2009dq} was adapted to 
moving punctures and evolved in~\cite{Reifenberger2012}. 
However, only the lower-frequency $\left(2,2\right)$ mode of the 
junk radiation was studied.
We also point out that all the previous initial data sets using 
PN corrections only approximately satisfied the Einstein constraint 
equations, and were not used to provide free data for a constraint solver.
This had adverse manifestations in an evolution, such as the black 
holes losing mass.

The present work examines the effects of including realistic tidal
deformations in the context of superposed black-hole initial data,
both elucidating the origin and quantifying the amount of junk radiation 
that is associated with modeling the horizon geometries.
It also represents a first step in using PN results to construct
constraint-satisfying initial data that future efforts can build on.
In particular, we construct excision initial data for an equal-mass,
nonspinning black-hole binary in the
extended conformal-thin-sandwich formalism using a
similar method as in~\cite{Lovelace2008,Lovelace2009},  
by superposing two tidally perturbed Schwarzschild metrics given
in~\cite{JohnsonMcDaniel:2009dq}. The Einstein constraint equations are 
solved with the pseudospectral elliptic solver of~\cite{Pfeiffer2003}, 
and the initial data are evolved for the early inspiral phase 
following the techniques found in~\cite{Szilagyi:2009qz}. 
These results are then contrasted with those for conformally flat initial 
data and superposed Kerr-Schild initial data, which both do not have 
realistic tidal deformations. 
 
This paper is organized as follows. Section~\ref{sec:IDformalism} 
summarizes the extended-conformal-thin-sandwich formalism for constructing 
initial data, and details our choices for the freely specifiable data and 
boundary conditions. 
Section~\ref{sec:Ev} presents the evolutions of our initial data, and 
inspects various properties of the black hole horizons at early times. 
The junk radiation content that is generated in the evolutions is also 
quantified. 
Section~\ref{sec:Discussion} gives final remarks on 
our results and discusses potential directions for future work.

%##############################################################################
\section{Initial Data}
\label{sec:IDformalism}
%##############################################################################
\subsection{Extended-conformal-thin-sandwich equations}
Initial data is constructed within the extended-conformal-thin-sandwich
formalism~\cite{York1999,Pfeiffer2003b}.
First, the spacetime metric is decomposed into $3+1$ form~\cite{ADM,york79}
\begin{align}
\label{eq:3Plus1Metric}
^{(4)}ds^2 &= g_{\mu\nu}dx^{\mu}dx^{\nu},\\
      &= -N^2dt^2+g_{ij}\left(dx^i+\beta^idt\right)\left(dx^j+\beta^jdt\right),
\end{align}
where $g_{ij}$ is the spatial metric of a $t=\text{constant}$ hypersurface
$\Sigma_t$, $N$ is the lapse function, and $\beta^i$ is the shift vector.
(Here and throughout this paper, Greek indices are spacetime indices running
from 0 to 3, while Latin indices are spatial indices running from 1 to 3.)
The Einstein equations then become a set of evolution equations,
\begin{align} 
\label{eq:EvMetric}
(\partial_{t}-\mathcal{L}_{\beta})g_{ij} &= -2NK_{ij},\\
\label{eq:EvK}
(\partial_{t}-\mathcal{L}_{\beta})K_{ij} &= N\left(R_{ij}-2K_{ik}{K^k}_j+KK_{ij}\right)-
\nabla_i\nabla_jN,
\end{align}
and a set of constraint equations,
\begin{align}
\label{eq:Ham}
\SRicciS + \TrExCurv^2 - \ExCurv_{ij}\ExCurv^{ij} & = 0,\\
\label{eq:Mom}
\nabla_j\left(\ExCurv^{ij}-\SMetric^{ij}\TrExCurv\right) & = 0.
\end{align}
Equation~\eqref{eq:Ham} is known as the Hamiltonian constraint,
and Eq.~\eqref{eq:Mom} is the momentum constraint.
In the above, all matter source terms have been neglected, since we will
only be interested in vacuum spacetimes. Also, $\mathcal{L}$ is the
Lie derivative, $\nabla_i$ is the covariant derivative compatible
with $\SMetric_{ij}$, $\SRicciS=\SMetric^{ij}\SRicci_{ij}$ is the trace of the
Ricci tensor $\SRicci_{ij}$ of $\SMetric_{ij}$, and
$K=\SMetric^{ij}\ExCurv_{ij}$ is the trace of the extrinsic curvature
$\ExCurv_{ij}$ of $\Sigma_t$.

\begin{table*}[!ht]
%\begin{tabular}{|p{2cm} p{2cm} p{2cm} p{1.5cm} p{2cm} p{2cm} p{2cm} p{2cm}|}
\begin{tabular}{>{\centering\arraybackslash}m{2cm} | >{\centering\arraybackslash}m{1.9cm} >{\centering\arraybackslash}m{1.9cm} >{\centering\arraybackslash}m{1.4cm} >{\centering\arraybackslash}m{1.4cm} >{\centering\arraybackslash}m{1.8cm} >{\centering\arraybackslash}m{2cm} >{\centering\arraybackslash}m{1.8cm} >{\centering\arraybackslash}m{2.5cm} >{\centering\arraybackslash}m{0.000001cm}}
\hline
{\bf Initial data} & $\boldsymbol{M_{\rm ADM}/M}$ & $\boldsymbol{J_{\rm ADM}/M^2}$ & $\boldsymbol{d_0/M}$ & $\boldsymbol{s_0/M}$ & $\boldsymbol{M\Omega_0}$ & $\boldsymbol{\dot{r}_0/r_0}$ & $\boldsymbol{e}$ & $\boldsymbol{\chi_0}$ & \\[0.1cm]
\hline
CFMS & 0.992402 & 1.0898 & 14.44 & 17.37 & 0.0167081 & $-2.84\times10^{-5}$ & $5\times10^{-5}$ & $2.05574\times10^{-7}$ & \\[0.1cm]
SKS  & 0.992629 & 1.1046 & 15.0 & 16.90 & 0.0158313 & $-4.79\times10^{-5}$ & $2\times10^{-4}$ & $1.05715\times10^{-7}$ & \\[0.1cm]
STPv1  & 0.992699 & 1.1223 & 15.0 & 17.68 & 0.0156354 & 0.0 & $1\times10^{-2}$ & $4.33909\times10^{-6}$ \\[0.1cm]
STPv2  & 0.992539 & 1.1093 & 15.0 & 17.68 & 0.0156354 & 0.0 & $1\times10^{-2}$ & $3.51594\times10^{-5}$ \\[0.1cm]
\hline
\end{tabular}
\caption{
Summary of our initial data properties, where $M_\text{ADM}$ is the
Arnowitt-Deser-Misner (ADM) energy, $J_\text{ADM}$ is the
ADM angular momentum, $d_0$ is the
initial coordinate separation, $s_0$ is the initial proper separation,
$\Omega_0$ is the initial orbital frequency, $\dot{r_0}/r_0$ is the
initial radial velocity of one black hole, $e$ is the eccentricity,
and $\chi_0$ is the initial dimensionless spin magnitude of one black hole.
\label{tab:IDTable}}
\end{table*}

The spatial metric is decomposed in terms of a conformal metric
$\CMetric_{ij}$ and a conformal factor $\CF$,
\begin{equation}
\label{eq:SMetric-ID}
g_{ij}=\CF^4\CMetric_{ij}.
\end{equation}
The tracefree time derivative of the conformal metric is denoted by
\begin{equation}
\tilde{u}_{ij}=\partial_{t}\CMetric_{ij},
\end{equation}
and satisfies $\tilde{u}_{ij}\tilde{g}^{ij}=0$.
A conformal lapse is also defined by $\tilde{N}=\CF^{-6}N$.
Treating $N\psi=\tilde{N}\psi^7$ as an independent variable, 
Eqs.~\eqref{eq:Ham},~\eqref{eq:Mom}, and the trace
of~\eqref{eq:EvK} can then be written as
\begin{align}
\label{eq:XCTS-Ham}
\tilde{\nabla^2}\psi-\frac{1}{8}\psi\tilde{R}-\frac{1}{12}\psi^5K^2+
\frac{1}{8}\psi^{-7}\tilde{A}_{ij}\tilde{A}^{ij} = 0,\\
\label{eq:XCTS-Mom}
\tilde{\nabla}_j\left(\frac{\psi^7}{2\left(N\psi\right)}%
\left(\mathbb{L}\beta\right)^{ij}\right)-\tilde{\nabla}_j
\left(\frac{\psi^7}{2\left(N\psi\right)}\tilde{u}^{ij}\right)%
-\frac{2}{3}\psi^6\tilde{\nabla}^iK = 0,\\
\label{eq:XCTS-EvK}
\tilde{\nabla}^2\left(N\psi\right)%
-\left(N\psi\right)\left(\frac{1}{8}\tilde{R}+
\frac{5}{12}\psi^4K^2+\frac{7}{8}\psi^{-8}\tilde{A}_{ij}\tilde{A}^{ij}\right)%
\notag\\
= -\psi^5\left(\partial_tK-\beta^k\partial_kK\right).
\end{align}
In the above, $\tilde{\nabla}_i$ is the covariant derivative compatible with
$\tilde{g}_{ij}$, $\tilde{R}=\tilde{g}^{ij}\tilde{R}_{ij}$ is the trace of the
Ricci tensor $\tilde{R}_{ij}$ of $\tilde{g}_{ij}$, $\tilde{\mathbb{L}}$
is the longitudinal operator,
\begin{equation}
\left(\tilde{\mathbb{L}}\beta\right)^{ij}=%
\tilde{\nabla}^i\beta^j+\tilde{\nabla}^j\beta^i-
\frac{2}{3}\tilde{g}^{ij}\tilde{\nabla}_k\beta^k,
\end{equation}
and $\tilde{A}^{ij}$ is
\begin{equation}
\tilde{A}^{ij}=\frac{1}{2\tilde{N}}%
\left(\left(\tilde{\mathbb{L}}\beta\right)^{ij}-
\tilde{u}^{ij}\right),
\end{equation}
which is related to $K_{ij}$ by
\begin{equation}
\label{eq:K-ID}
K_{ij}=\psi^{-10}\tilde{A}_{ij}+\frac{1}{3}g_{ij}K.
\end{equation}

Given a particular choice for the freely specifiable data 
\begin{equation}
\label{eq:freedata}
\left(\tilde{g}_{ij}, \tilde{u}_{ij}, K, \partial_tK\right), 
\end{equation}
Eqs.~\eqref{eq:XCTS-Ham},~\eqref{eq:XCTS-Mom}, 
and~\eqref{eq:XCTS-EvK} constitute a coupled set of elliptic equations 
that one solves for $\psi$, $N\psi$, and $\beta^i$.
These equations are known as the 
extended-conformal-thin-sandwich equations.
From their solutions, the physical initial data $g_{ij}$ and
$K_{ij}$ are obtained from Eqs.~\eqref{eq:SMetric-ID} and~\eqref{eq:K-ID},
respectively.

\begin{figure}[t]
\includegraphics[scale=0.5]{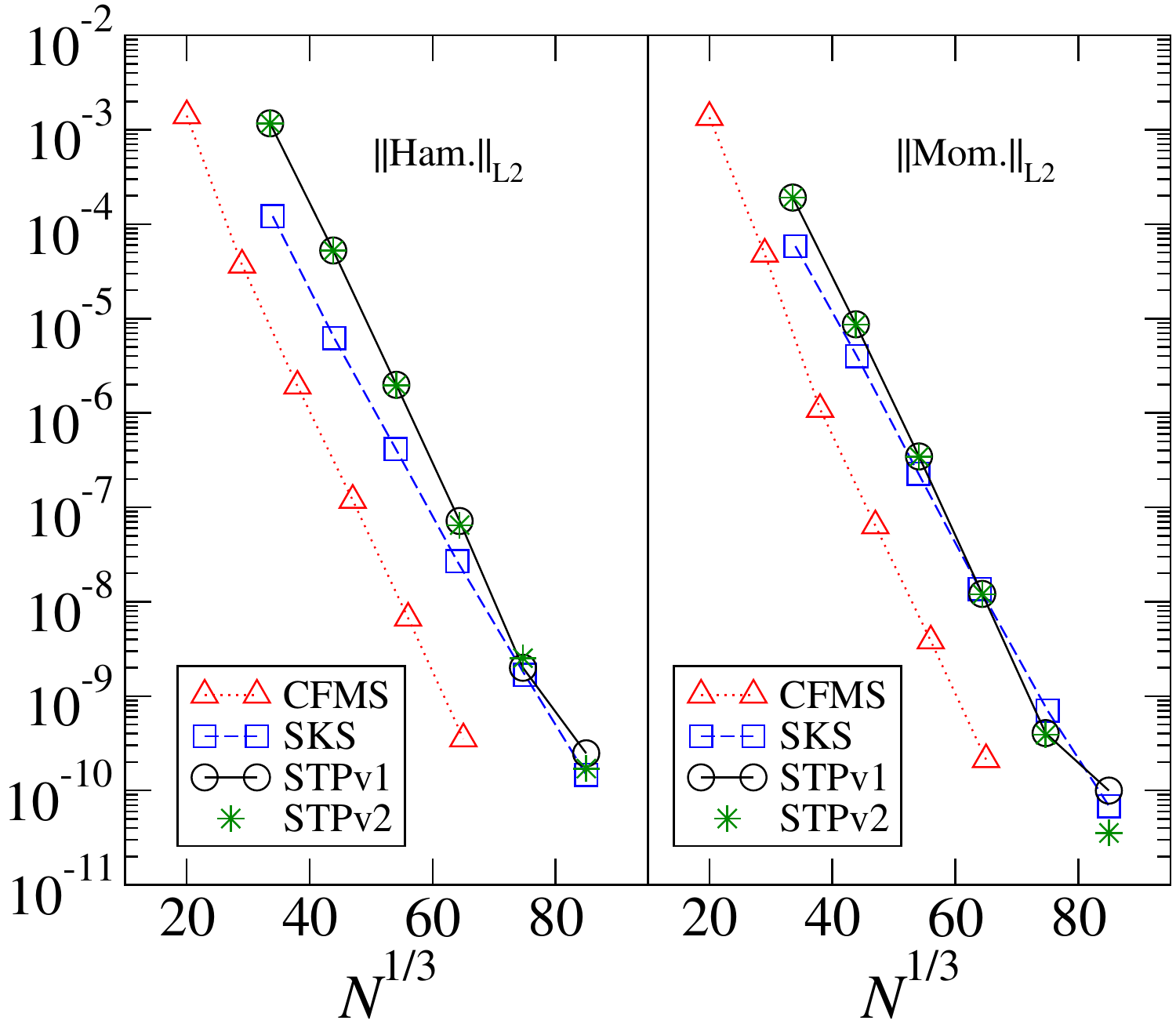}
\caption{
\label{fig:ConstraintsID_All_Smaller}
Convergence of the $L^2$ norms of the Hamiltonian and momentum constraints for
different types of initial data, with increasing number of
grid points $N$ in the computational domain.
}
\end{figure}

%------------------------------------
\subsection{Types of initial data}
\label{sec:ID_Types} 
%------------------------------------

Our initial data represent two equal-mass, nonspinning black holes, 
each with an initial Christodoulou mass $m=M/2$, which are situated 
$\sim$16 orbits before merger. 
Equations ~\eqref{eq:XCTS-Ham},~\eqref{eq:XCTS-Mom}, 
and~\eqref{eq:XCTS-EvK} are solved with the pseudospectral 
elliptic solver detailed in~\cite{Pfeiffer2003}. 
The singularities of the black holes are excised from the 
computational domain. 
The initial data sets described below differ in their choices for 
the freely specifiable $\tilde{g}_{ij}$ and $K$, and the boundary 
conditions imposed at the excision surfaces $\mathcal{S}$. 
In all cases, we set the freely specifiable time derivatives 
to zero in the corotating frame of the binary,
\begin{equation}
\tilde{u}_{ij}=0,\hspace{2 mm}\partial_t K=0.
\end{equation}
In terms of a radial coordinate $r$ measured from the 
center of mass, the outer boundary is placed at a large distance
$R_{\text{large}}/M=1.0\times10^9$, 
where asymptotic flatness is imposed in the inertial frame of the binary,
\begin{align}
\left.\psi\right|_{R_{\text{large}}}&=N\psi\vert_{R_{\text{large}}}=1, \\
%\tilde{N}\vert_{R_{\text{outer}}}&=1, \\
\left.\beta^i\right|_{R_{\text{large}}}&=0.
\end{align}

%-------------------------------------------
\subsubsection{Conformally flat, maximally sliced initial data}
\label{sec:CFMS_ID}
%-------------------------------------------
%The simplest type of initial data that we consider is 
%conformally flat and maximally sliced (CFMS), identical 
%to that used in~\cite{Boyle2007,Scheel2009}. 
Conformal flatness is currently the most common choice in 
constructing black hole initial data, especially in the Bowen-York 
formulation~\cite{Bowen-York:1980} for codes that use puncture methods, 
because the constraint equations, Eqs.~\eqref{eq:Ham}--\eqref{eq:Mom}, 
simplify greatly. 
For our purposes, we consider conformally flat and maximally sliced (CFMS) 
initial data, identical to that used in~\cite{Boyle2007,Scheel2009}.
That is, the remaining free data are fixed by a flat conformal metric,
\begin{equation}
\tilde{g}_{ij} = \delta_{ij},
\end{equation}
along with the maximal slicing condition~\cite{Lichnerowicz44,Smarr78b},
\begin{equation}
K=0.
\end{equation}
Quasi-equilibrium boundary conditions~\cite{Cook2004} for 
$\psi$ and $\beta^i$ are imposed on spherical excision surfaces,
with $\left.\beta^i\right|_{\mathcal{S}}$ adjusted so that the spins 
$\chi$~\cite{Lovelace2008} of the black holes are very small.
The lapse boundary condition is given by~\cite{Caudill-etal:2006} 
\begin{equation}
\frac{d\left(N\psi\right)}{dr}\bigg|_\mathcal{S}=0.
\end{equation}

The initial orbital eccentricity was reduced to 
$e\sim5\times10^{-5}$ using the iterative procedure 
in~\cite{Pfeiffer-Brown-etal:2007}, giving an initial orbital 
frequency $M\Omega_0=0.0167081$ and radial velocities 
$\dot{r}_0/r_0=-2.84\times10^{-5}$. 
Properties of CFMS initial data are summarized in 
Table~\ref{tab:IDTable}. 
The convergence of the Hamiltonian and momentum constraints
are shown in Fig.~\ref{fig:ConstraintsID_All_Smaller}, as the red triangles
connected by the dotted lines.
Plotted is the $L^2$ norm\footnotemark of the Hamiltonian constraint
and the root-sum-square of the $L^2$ norms of the momentum constraint
components, versus the total number of grid points $N$ in the
initial data domain. Due to the simple structure of the conformal 
data, less grid points are needed to resolve the solution in contrast  
to the other types of conformally curved initial data we present below.

\footnotetext[1]{
The $L^2$ norm of a tensor $T_{ijk\cdots}\left(x\right)$ evaluated at
$N$ grid points $x_i$ is defined as
\begin{equation}
\left|\left|T_{ijk\cdots}\right|\right|_{L^2}:=\sqrt{\frac{1}{N}\displaystyle\sum_{i=0}^N\bar{T}^2\left(x_i\right)},
\end{equation}
where
\begin{equation}
\bar{T}^2:=T_{ijk\cdots}T_{i^\prime j^\prime k^\prime\cdots}\delta^{ii^\prime}\delta^{jj^\prime}\delta^{kk^\prime}\cdots.
\end{equation}
}

%While these choices simplify the constraint equations considerably, 
%and maximal slicing has been shown to be useful in numerical 
%simulations due to its singularity-avoidance 
%property~\cite{smarr_etal76,Naka81,StPi85,evans86}, 
%a flat conformal metric is not physically well-justified and 
%is a main source of the junk radiation that is generated during 
%an evolution. It is then desirable to seek alternative initial data 
%that is better motivated.

\footnotetext[2]{
This type of initial data is particularly useful for simulating 
highly spinning black holes~\cite{Lovelace:2011nu}, 
although we do not make use of that aspect here.
}

\subsubsection{Superposed Kerr-Schild initial data}
\label{sec:SKS_ID}
Another type of initial data in use is based on the superposition of 
black hole metrics, as first presented 
in~\cite{Matzner1999,Marronetti-Matzner:2000,Marronetti2000}.
Here we follow the variant of Lovelace \textit{et al.}~\cite{Lovelace2008} 
in constructing superposed Kerr-Schild (SKS) initial data.\footnotemark~ 
In this approach, the remaining free data are taken 
to be a weighted superposition of the corresponding quantities for 
a boosted, nonspinning Kerr-Schild black hole, 
%conformal metric is taken to be 
%a weighted superposition of two Kerr-Schild metrics,
\begin{align}
\label{eq:SuperposedMetric}
\tilde{g}_{ij}&=\delta_{ij}+\displaystyle\sum\limits_{a=1}^2e^{-r^2_a/w^2_a}\left(g^a_{ij}-\delta_{ij}\right), \\
\label{eq:SuperposedK}
K&=\displaystyle\sum\limits_{a=1}^2e^{-r^2_a/w^2_a}K_a,
\end{align}
where $g^a_{ij}$ and $K_a$ are the spatial metric and trace of the 
extrinsic curvature, respectively, of a Kerr-Schild black hole 
(labeled by $a$) with mass 
$\tilde{m}^{\text{KS}}_a$ and speed $\tilde{v}^{\text{KS}}_a$.
The Gaussian factor $e^{-r^2_a/w^2_a}$, for a fixed weight parameter 
$w_a$, is a function of Euclidean distance $r_a$ from black hole $a$.
Its presence ensures that in the vicinity of each black hole, 
the conformal data approach the appropriate Kerr-Schild values, 
while far away from the black holes the conformal data approach that of 
a flat spacetime.

Unlike CFMS initial data, the excision surfaces  
are not coordinate spheres, but are Lorentz-contracted along the 
direction of the boost. 
Quasi-equilibrium boundary conditions~\cite{Cook2004} for 
$\psi$ and $\beta^i$ are still imposed on $\mathcal{S}$, with the 
black hole spins $\chi$ made very small. 
However, the following Dirichlet boundary condition for the lapse is used,
\begin{equation}
\left.N\psi\right|_{\mathcal{S}}=1+\displaystyle\sum\limits_{a=1}^2%
e^{-r^2_a/w^2_a}\left(N_a-1\right),
\end{equation}
where $N_a$ is the lapse of the corresponding Kerr-Schild black hole.

For the parameters entering our SKS initial data, we use 
$\tilde{m}^{\text{KS}}_a=0.37298$, $w_a/\tilde{m}^{\text{KS}}_a=6$, and 
$\tilde{v}^{\text{KS}}_a=0.11873$.
The value of $\tilde{m}^{\text{KS}}_a$ was set so that $m=0.5$ 
and $M=1$. Also, the value of $w_a$ was chosen to approximately 
minimize the junk radiation content.
The convergence of the Hamiltonian and momentum constraints are shown 
in Fig.~\ref{fig:ConstraintsID_All_Smaller}, as the blue squares connected by 
the dashed lines.

The initial eccentricity was reduced to $e\sim2\times10^{-4}$ using the 
iterative procedure in~\cite{Pfeiffer-Brown-etal:2007}, 
giving an initial orbital frequency $M\Omega_0=0.0158313$ and 
radial velocities $\dot{r}_0/r_0=-4.79\times10^{-5}$. 
Although the eccentricity could have been reduced further without 
difficulty, doing so was not important for our present purposes. 
Properties of SKS initial data are summarized in Table~\ref{tab:IDTable}.
The convergence of the Hamiltonian and momentum constraints are shown
in Fig.~\ref{fig:ConstraintsID_All_Smaller}, as the blue squares connected by
the dashed lines.

%-------------------------------------------
\subsubsection{Superposed tidally perturbed initial data}
\label{sec:STP_ID}
%-------------------------------------------
To include realistic tidal deformations of the black holes in our 
initial data, we build on the method presented above and construct 
superposed tidally perturbed (STP) initial data. 
Instead of Kerr-Schild metrics, which only characterize 
isolated black holes, we make use of suitable 
tidally perturbed black hole metrics that have been determined by 
Johnson-McDaniel \textit{et al.}~\cite{JohnsonMcDaniel:2009dq}. 
%We briefly describe the results of their work that we make use of here.
Their metrics are obtained by asymptotically 
matching perturbed Schwarzschild metrics in horizon-penetrating 
Cook-Scheel harmonic coordinates~\cite{cook_scheel97}, 
to an order $\mathcal{O}\left(v^4\right)$ 
PN near zone metric~\cite{Blanchet-Faye-Ponsot:1998} 
in harmonic coordinates for two point particles in a circular orbit. 
To perform the matching, the black hole metrics are 
perturbatively transformed to the same coordinate system as the PN 
metric. The matching then determines this coordinate transformation up to 
$\mathcal{O}\left(v^5\right)$, as well as the Newtonian quadrupole 
and octupole tidal fields and the 1PN corrections to the quadrupole 
fields in the black hole metrics. 
In the appendix, we collect the explicit expressions for these quantities.

The fact that horizon-penetrating coordinates are used for the black 
hole metrics from the start is important for us, even though the PN 
harmonic coordinates themselves are not horizon-penetrating. This is 
because after perturbatively transforming to PN harmonic coordinates, 
the black hole metrics remain in horizon-penetrating coordinates, 
so that we can use excision in constructing initial data. 
Nevertheless, we only apply the transformation up to 
$\mathcal{O}\left(v^3\right)$, as the lowest-order piece between 
Cook-Scheel and PN harmonic coordinates for an unperturbed Schwarzschild 
black hole enters at $\mathcal{O}\left(v^4\right)$, and we find that the 
constraint violations (before solving the constraint equations) are 
larger when the higher-order pieces are included.
This should not cause too much concern for our present purposes, 
since we are not yet including the PN near zone metric in our conformal 
data, so that placing the black hole metrics as closely as possible 
into PN harmonic coordinates is not crucial. 
Moreover, the lowest-order pieces of the Lorentz boost for the 
black holes' orbital 
motion are already present at $\mathcal{O}\left(v^3\right)$, 
which is the highest order for which all pieces of the coordinate 
transformation are fixed by the matching.

The remaining free data for our STP initial data are then chosen as in 
Eqs.~\eqref{eq:SuperposedMetric}--\eqref{eq:SuperposedK}, 
but now $g^a_{ij}$ and $K_a$ are the 
spatial metric and trace of the extrinsic curvature, respectively, of 
a tidally perturbed black hole mentioned above.
Thus, while the conformal data include realistic tidal deformations in 
the vicinity of each black hole, they still approach that of a flat 
spacetime between and far away from the black holes. That is, there are 
no PN interaction terms in the near zone surrounding the black holes, 
nor any outgoing radiation content in the wave zone.

We use spherical excision surfaces, and impose the following 
Dirichlet boundary conditions there,
\begin{align} 
\left.N\psi\right|_{\mathcal{S}} &= 1+\sum_{a=1}^2e^{-r_a^2/w_a^2}(N_a-1), \\
\left.\beta^i\right|_{\mathcal{S}} &= \sum_{a=1}^2e^{-r_a^2/w_a^2}\beta_a^i,
\end{align}
where $N_a$ and $\beta_a^i$ are the lapse and shift, respectively, 
of the corresponding tidally perturbed black hole.
The boundary condition on $\beta^i$ also fixes the black hole spins $\chi_0$, 
which still have values close to zero (cf. Table~\ref{tab:IDTable}).
We do not use quasi-equilibrium boundary conditions~\cite{Cook2004} here, 
since they were derived by requiring an excision surface to be 
an apparent horizon on which the outgoing null normals have vanishing shear.
For a tidally perturbed black hole however, the apparent horizon foliates 
a dynamical horizon~\cite{Booth2005} and has non-vanishing shear, 
so imposing quasi-equilibrium boundary conditions would not be  
appropriate.
%The quasiequilibrium boundary conditions could be generalized to
%dynamical horizons, but we have not yet attempted to do so.
%Imposing the Dirichlet boundary conditions above does have an advantage
%though, which is that the apparent horizons in our initial data lie 
%outside the excision surfaces. This eliminates the need later on to 
%extrapolate the initial data onto an evolution grid with smaller 
%excision boundaries, as in the case of CFMS and SKS initial data, 
%which fills those subdomains inside (and possibly containing) 
%the horizons with constraint-violating data.

%In the appendix, we collect the explicit expressions 
%used for the tidally deformed black hole metric in Cook-Scheel 
%harmonic coordinates and the associated coordinate transformation to 
%PN harmonic coordinates. 

There are several additional parameters that we need to choose in 
constructing our initial data.
These are the mass parameter $\tilde{M}$ in Cook-Scheel coordinates of 
Eqs.~\eqref{eq:CSmetric}--\eqref{eq:CSmetricfunctions}, 
plus the mass parameters $\tilde{m}_1=\tilde{m}_2$ 
and separation parameter $b$ in PN harmonic coordinates that appear 
in Eqs.~\eqref{eq:TidalFields}--\eqref{eq:PNtoCS_F3}.
Since $\tilde{M}$ and $\tilde{m}_1=\tilde{m}_2$ are found to be 
identical up to the highest order fixed by the asymptotic 
matching~\cite{JohnsonMcDaniel:2009dq}, 
we set them to equal to each other, 
$\tilde{M}=\tilde{m}_1=\tilde{m}_2=0.5036$. 
This value was chosen so that $m=0.5$ and $M=1$.
The separation parameter is simply taken to be the coordinate distance 
between the two apparent horizons in our initial data, that is 
$b/M=d_0/M=15$. In addition, we set $w_a/\tilde{M}=13$ for the 
superposition to approximately minimize the junk radiation content.

We consider two versions of STP initial data. 
The first verion incorporates all the terms in the tidal 
fields~\eqref{eq:TidalFields}--\eqref{eq:TidalFieldsLast}, which we 
call STPv1. The second, STPv2, includes only the lowest-order 
piece of the tidal field, the Newtonian electric quadrupole, 
with a ``corotating" time dependence 
obtained via the substitutions
\begin{align}
\label{eq:xcorotating}
\hat{x}_i&\rightarrow\hat{x}_i\text{cos}\omega t + \hat{y}_i\text{sin}\omega t,\\
\label{eq:ycorotating}
\hat{y}_i&\rightarrow-\hat{x}_i\text{sin}\omega t + \hat{y}_i\text{cos}\omega t.
\end{align}
Having these versions allows us to investigate the importance of 
including the higher-order contributions of the tidal fields.
This is especially relevant given that recently a lowest-order 
tidally perturbed metric with spin, with a corotating time dependence of 
the tidal fields, has become available~\cite{Gallouin:2012}. 

It should be noted that the parameters $\tilde{m}_1$, $\tilde{m}_2$, 
and $b$ used in transforming from Cook-Scheel to PN harmonic 
coordinates also affect the initial boost velocities of the black holes. 
The black holes would be on very nearly circular orbits, if the initial data 
were exactly in PN harmonic coordinates 
(and assuming the black holes are sufficiently 
far apart so that the PN approximation is valid). 
Due to the superposition procedure, our initial data are not in 
PN harmonic coordinates, and this gives rise to a rather large 
initial eccentricity of $e$$\sim$0.01.
The eccentricity could possibly be reduced by adjusting the values of 
these parameters, and the weight parameters $w_a$ in the superposition, 
as the latter also influence the properties of the black holes. 
Another way is to include higher-order boost pieces in the coordinate 
transformation, with an adjustable free 
parameter~\cite{Nathan:Email:July27-2012}.
For this paper though, we do not explore the results of these procedures.

Properties of our STP initial data are summarized in Table~\ref{tab:IDTable}.
The convergence of the Hamiltonian and momentum constraints are shown
in Fig.~\ref{fig:ConstraintsID_All_Smaller}. The values of the constraints are 
displayed as the black circles connected by solid lines for STPv1 
initial data, and the unconnected green stars for STPv2 initial data.  

%##############################################################################
\section{Evolutions}
\label{sec:Ev}
%##############################################################################

The initial data sets are evolved with similar methods as 
in~\cite{Szilagyi:2009qz}. We use the Spectral Einstein Code
({\tt SpEC})~\cite{Scheel2006,SpECwebsite} to solve a first-order 
representation~\cite{Lindblom2006} of the generalized
harmonic system~\cite{Friedrich1985,Garfinkle2002,Pretorius2005c}, on a 
computational domain from which the singularities are excised.
To accommodate the use of excision, two distinct coordinate frames are 
used: the ``grid frame'' that follows the motion of the black holes, 
and the ``inertial frame'' that is non-rotating and asymptotically 
Minkowski. 
The dynamical fields in the evolution equations are solved for in  
inertial-frame coordinates $x^i$, as functions of the grid-frame 
coordinates $\bar{x}^i$.

\begin{figure}[t]
\includegraphics[scale=0.5]{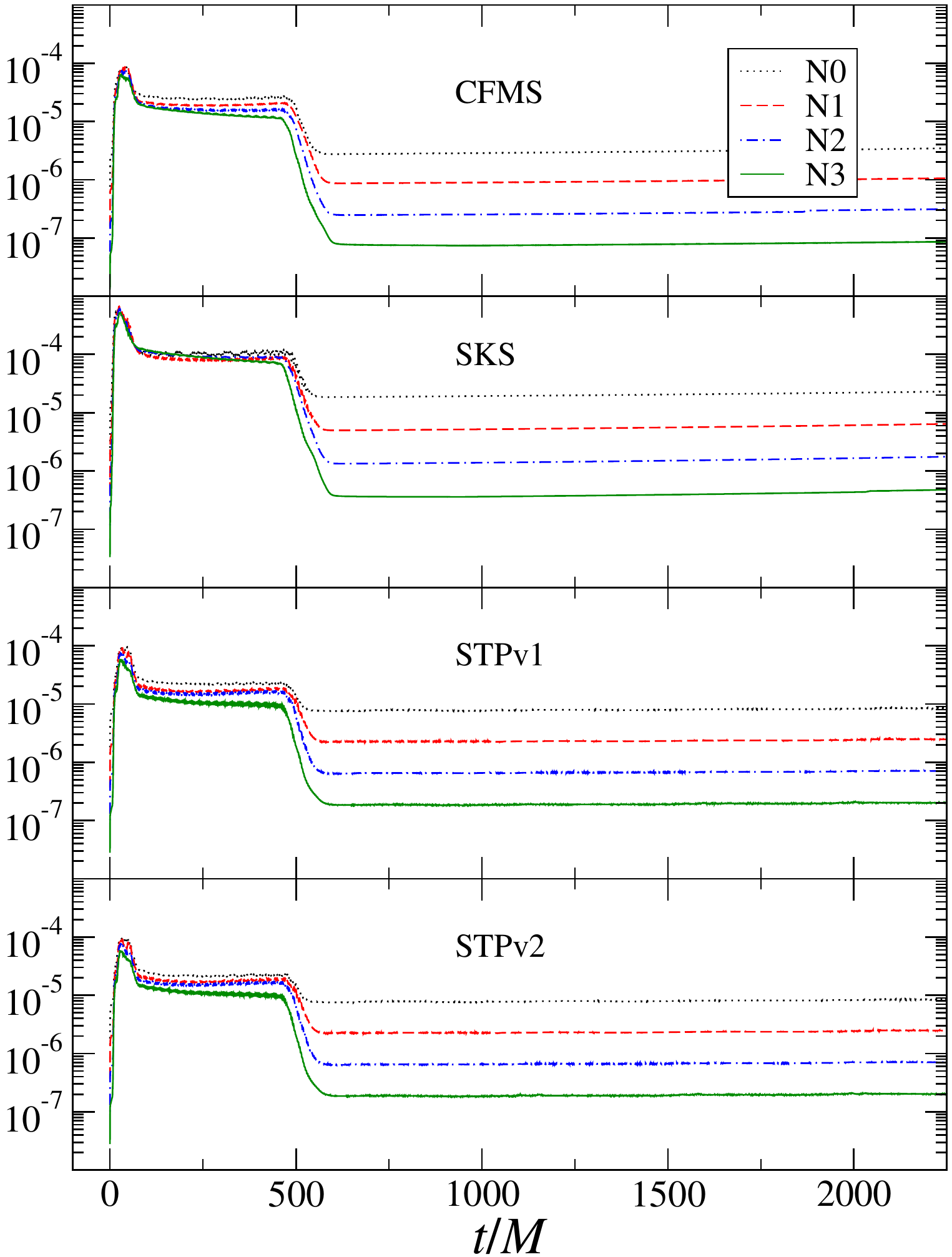}
\caption{
\label{fig:ConstraintsEv}
Constraint violations during the evolutions of different initial data.
Plotted is the $L^2$ norm of all constraints, normalized by the $L^2$
norm of the spatial gradients of all dynamical fields.
}
\end{figure}

In the computational domain, the excision boundaries are located just 
inside the apparent horizons, which differ marginally for the various types 
of initial data. The rest of the grid structures (consisting of 
touching spherical shells, cubes, and cylinders) remain the same. 
The outer boundaries are placed at $R_{\text{outer}}/M=480$. 
No boundary conditions are imposed at the excision surfaces, 
because all characteristic fields of the system are outgoing 
(into the black hole) there. The outer boundary 
conditions~\cite{Lindblom2006,Rinne2006,Rinne2007} imposed 
prevent the influx of unphysical constraint
violations~\cite{Stewart1998,FriedrichNagy1999,Bardeen2002,Szilagyi2002,%
Calabrese2003,Szilagyi2003,Kidder2005} and undesired incoming 
gravitational radiation~\cite{Buchman2006,Buchman2007}, 
while allowing the outgoing gravitational radiation to pass freely 
through. Interdomain boundary conditions are enforced with a penalty
method~\cite{Gottlieb2001,Hesthaven2000}.

The gauge freedom in the generalized harmonic system is fixed
through a freely specifiable gauge source function $H_{\mu}$ given by 
\begin{equation}
  H_{\mu}(t,x) = g_{\mu\nu}\nabla_{\lambda}\nabla^{\lambda}x^{\nu}
               = -\Gamma_{\mu},
\end{equation}
where $\Gamma_{\mu}=g^{\nu\lambda}\Gamma_{\mu\nu\lambda}$ is the trace of
the Christoffel symbol. In $3+1$ form, this becomes a set of 
evolution equations for $N$ and $\beta^i$~\cite{Lindblom2007}.
During the early inspiral, which is the only phase 
of the evolution considered here, the gauge is fixed by the 
quasi-equilibrium condition
\begin{equation}
\partial_{\bar{t}}\tilde{H}_{\mu}=0,
\end{equation}
where $\tilde{H}_{\mu}$ is a tensor defined such that 
$\tilde{H}_{\mu}=H_{\mu}$ in the inertial frame.

%In addition, we employ spectral adaptive mesh 
%refinement~\cite{Lovelace2010} during the evolutions, 
%in the subdomains close to and containing the black holes. 
%Depending on the truncation error requirement of the evolved 
%fields and the resolution requirement of the apparent horizons, 
%spectral basis functions are added or removed. The maximum truncation 
%error allowed for the evolved fields is set to $\tau=0.0001e^{-k}$, 
%where a different value of $k\in\left\{{0,1,2,3}\right\}$ 
%for each evolution determines its numerical resolution (low to high).

\begin{figure}[t]
\includegraphics[scale=0.5]{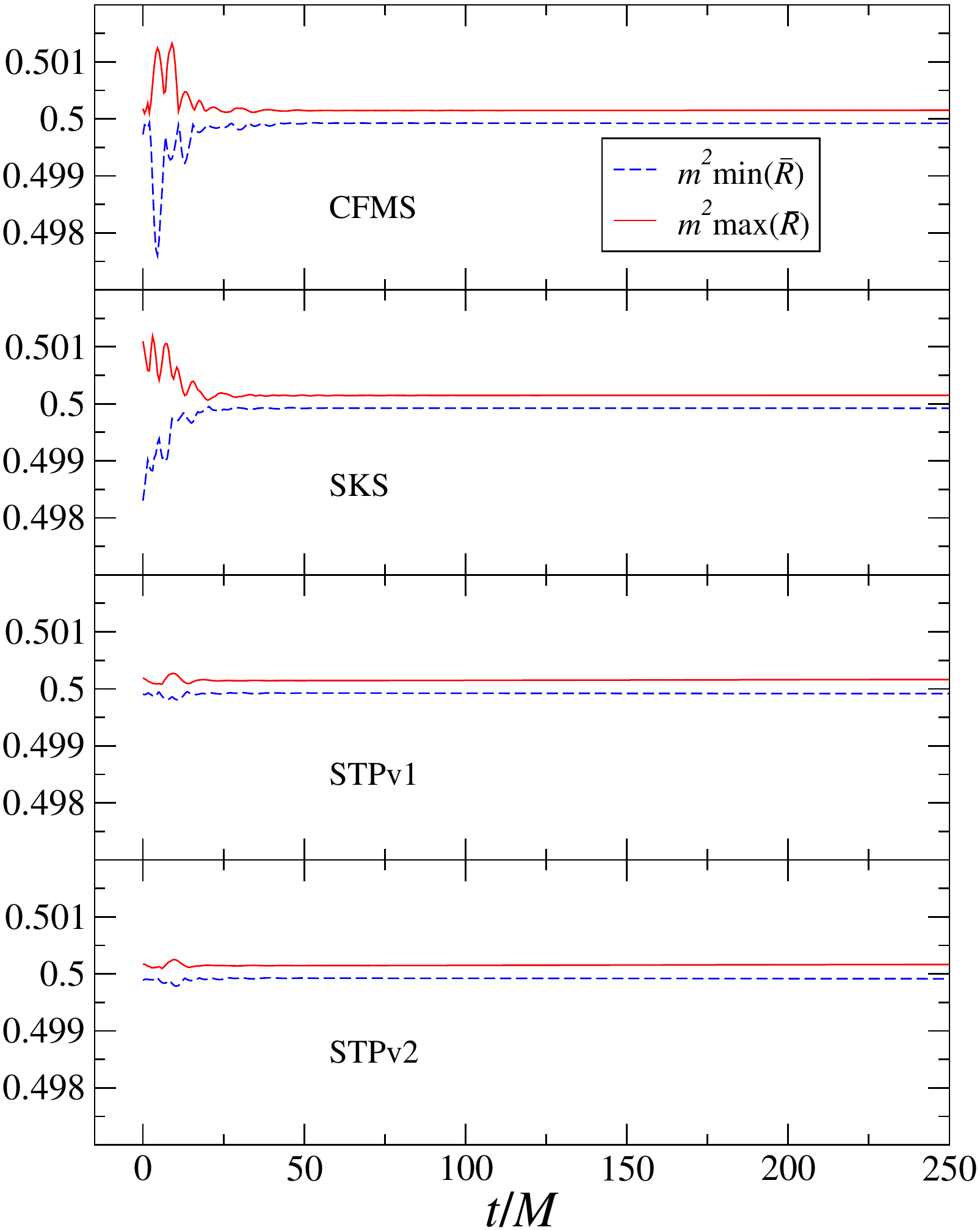}
\caption{
\label{fig:HorizonCurvatureComparison}
Extrema of the intrinsic scalar curvature $m^2\bar{R}$ over a black hole's
apparent horizon, in the evolution of different initial data sets.
}
\end{figure}

\begin{figure*}[h]
\centerline{
\includegraphics[width=0.35\textwidth, trim=0cm 3cm 0cm 0cm]{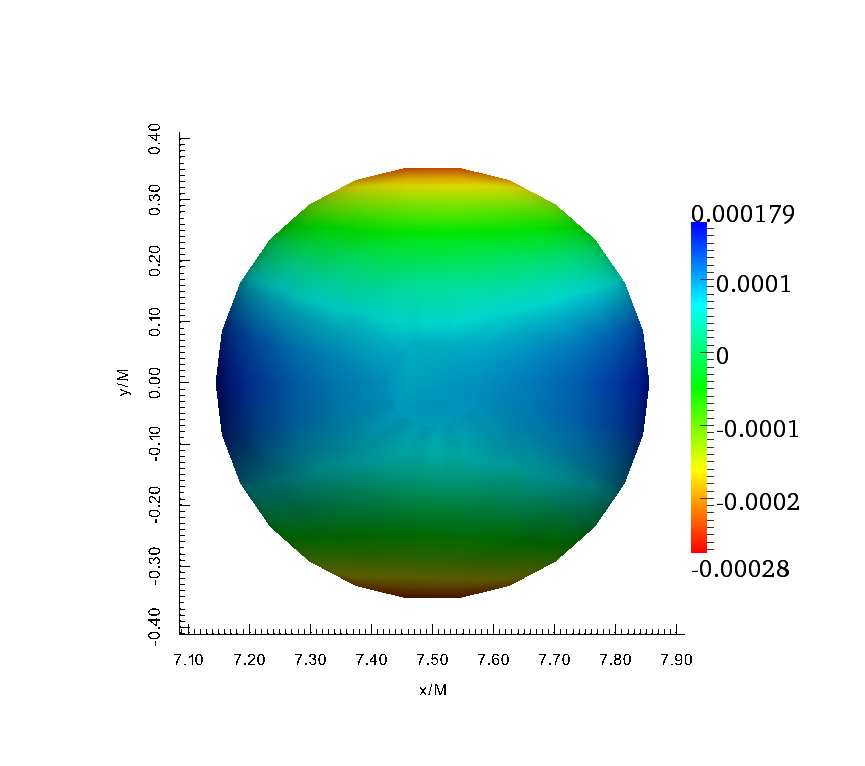}
$\qquad$\includegraphics[width=0.35\textwidth, trim=0cm 3cm 0cm 0cm]{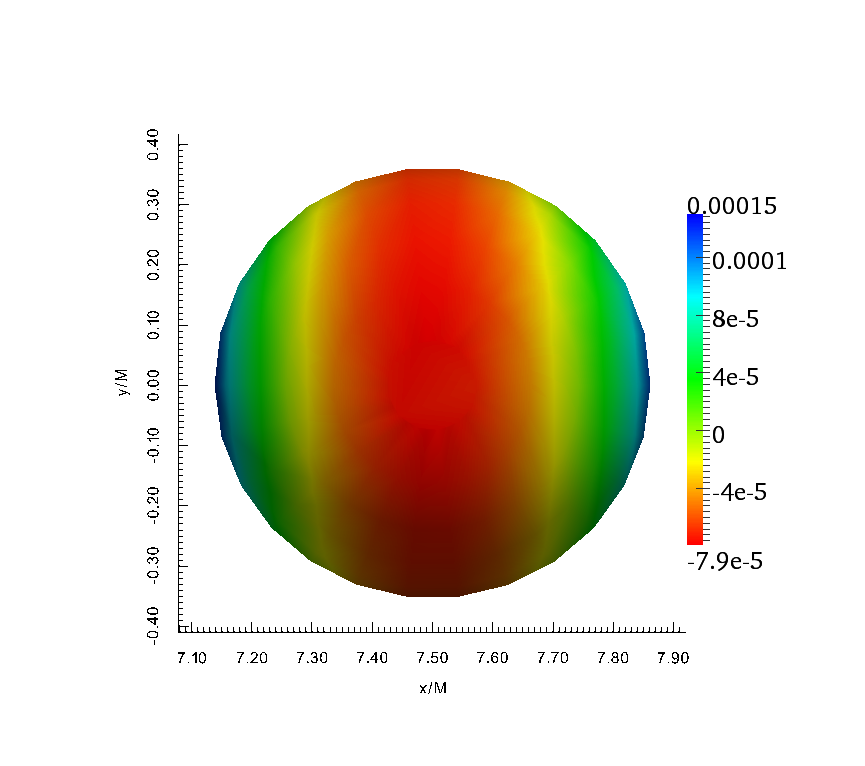}}
\caption{
\label{fig:HorizonComparisonCFMS}
Intrinsic scalar curvature $m^2\bar{R}$ of a black hole horizon in the
evolution of CFMS initial data, minus the Schwarzschild value of 0.5.
On the left are values at $t/M=0$.
On the right are values at $t_\text{relaxed}/M=250$.
}

\vspace{0.0mm}

\centerline{\includegraphics[width=0.35\textwidth, trim=0cm 3cm 0cm 0cm]{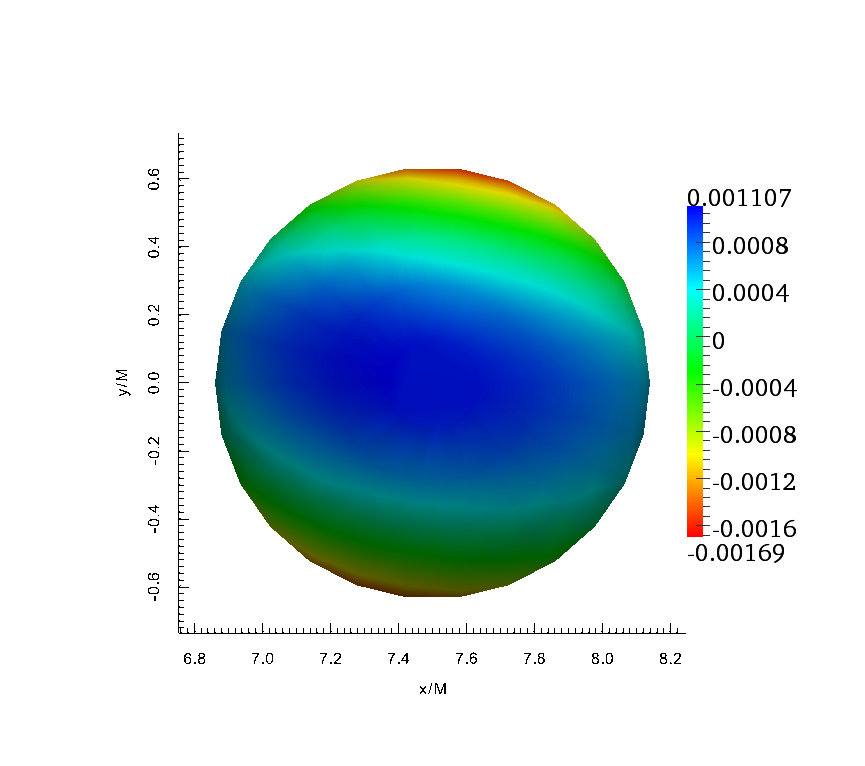}
$\qquad$\includegraphics[width=0.35\textwidth, trim=0cm 3cm 0cm 0cm]{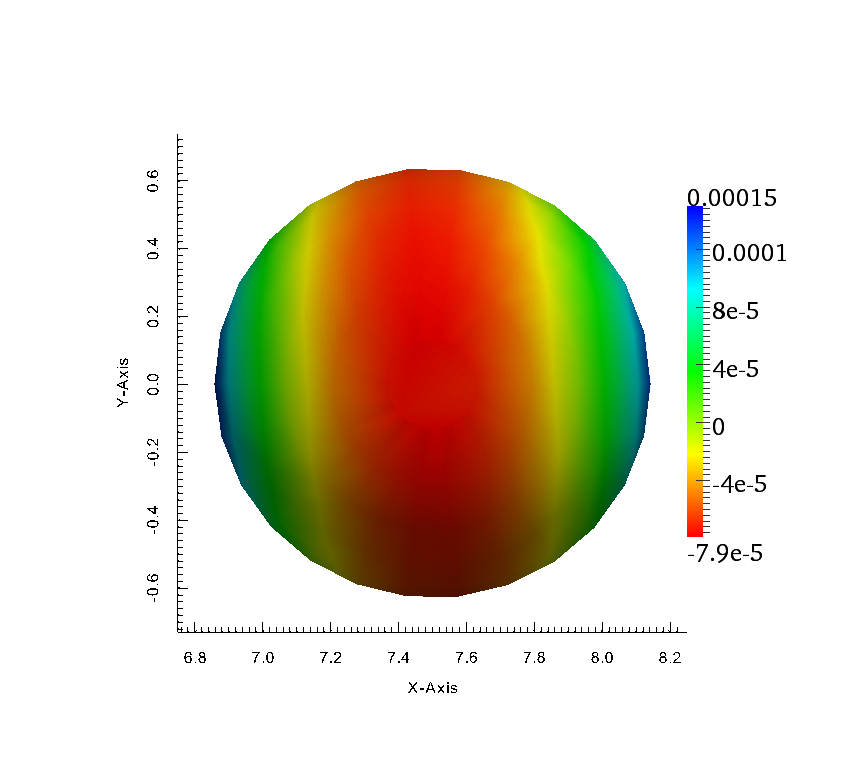}}
\caption{
\label{fig:HorizonComparisonSKS}
Intrinsic scalar curvature $m^2\bar{R}$ of a black hole horizon in the
evolution of SKS initial data, minus the Schwarzschild value of 0.5.
On the left are values at $t/M=0$.
On the right are values at $t_\text{relaxed}/M=250$.
}

\vspace{0.0mm}

\centerline{\includegraphics[width=0.35\textwidth, trim=0cm 3cm 0cm 0cm]{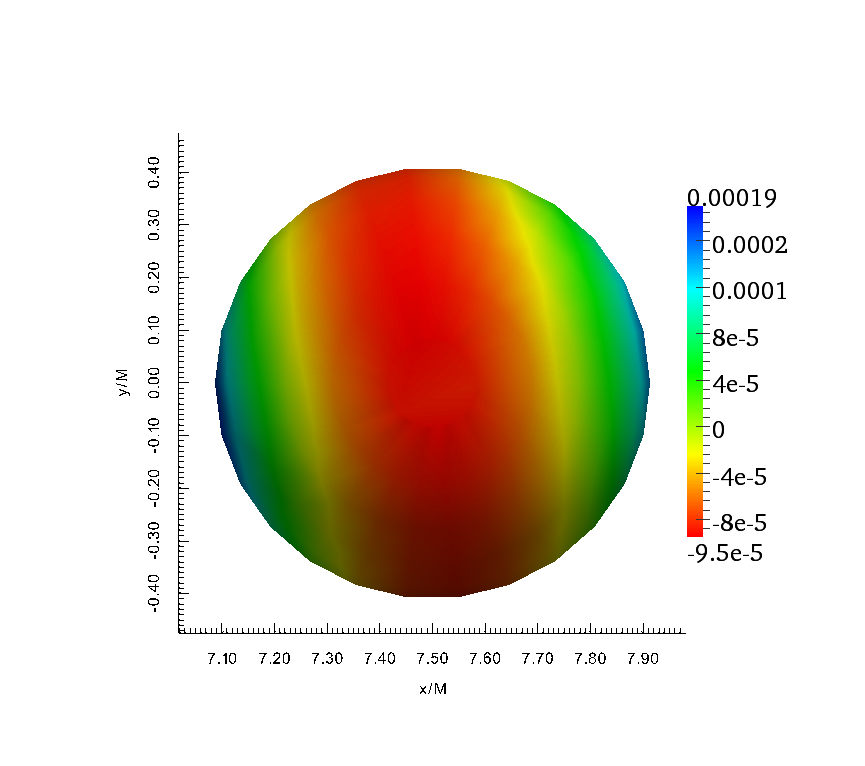}
$\qquad$\includegraphics[width=0.35\textwidth, trim=0cm 3cm 0cm 0cm]{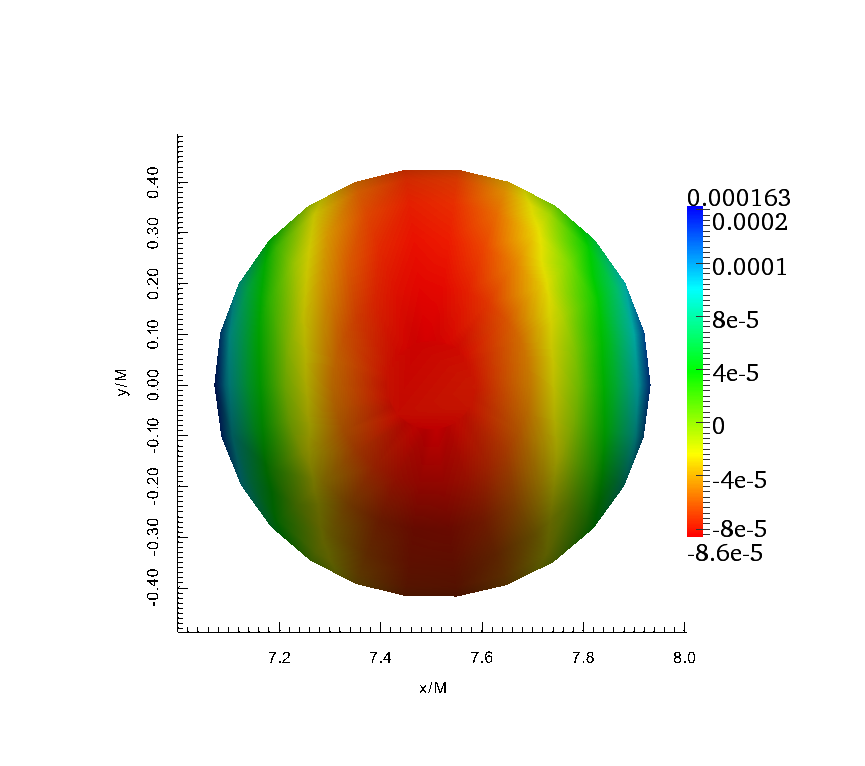}}
\caption{
\label{fig:HorizonComparisonSTPv1}
Intrinsic scalar curvature $m^2\bar{R}$ of a black hole horizon in the
evolution of STPv1 initial data, minus the Schwarzschild value of 0.5.
On the left are values at $t/M=0$.
On the right are values at $t_\text{relaxed}/M=250$.
}

\vspace{0.0mm}

\centerline{\includegraphics[width=0.35\textwidth, trim=0cm 3cm 0cm 0cm]{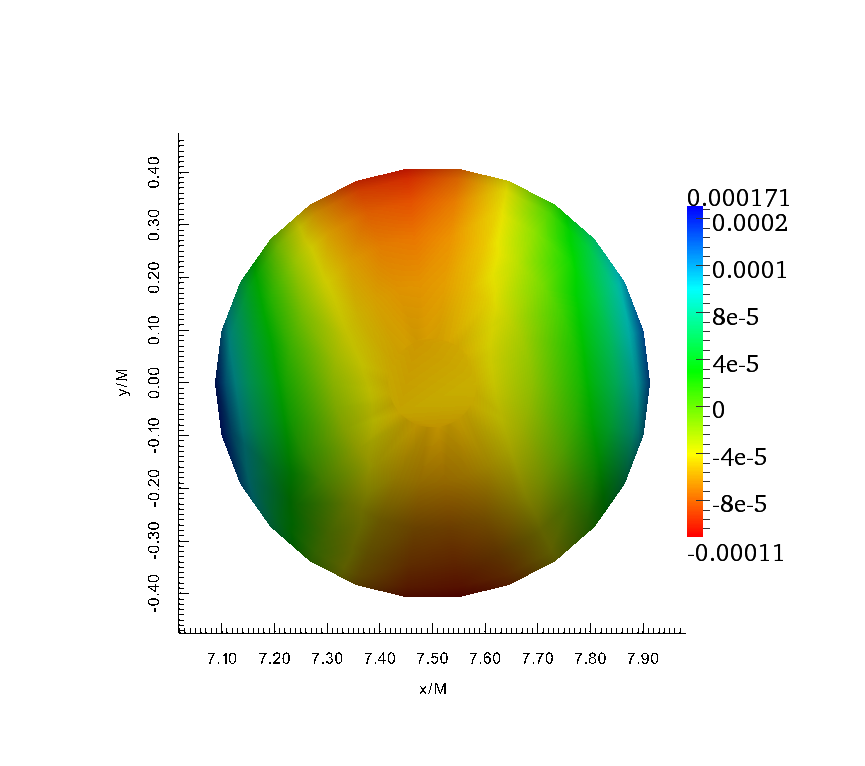}
$\qquad$\includegraphics[width=0.35\textwidth, trim=0cm 3cm 0cm 0cm]{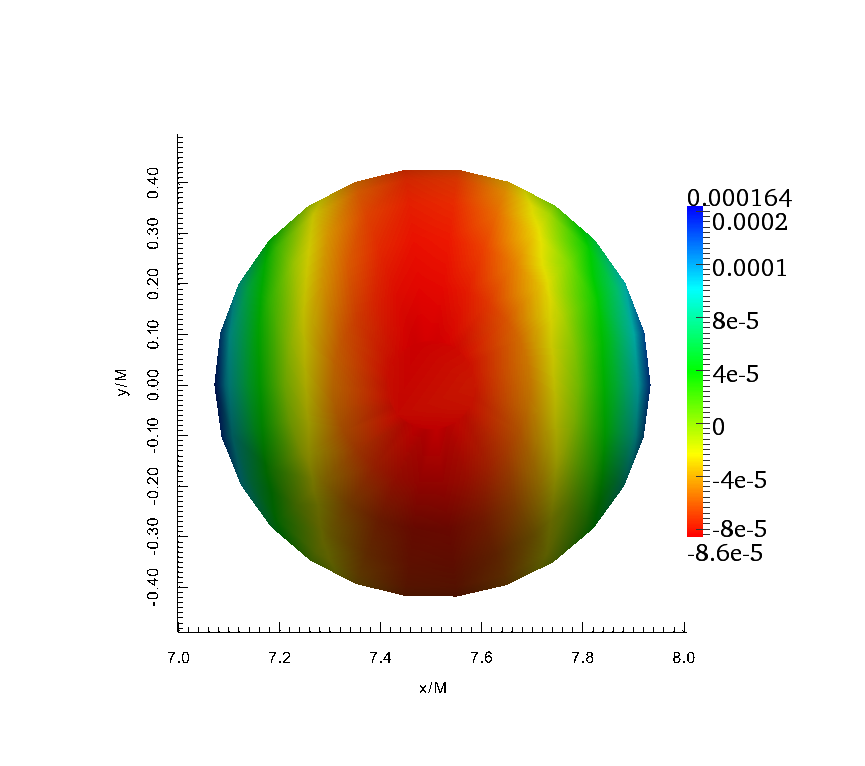}}
\caption{
\label{fig:HorizonComparisonSTPv2}
Intrinsic scalar curvature $m^2\bar{R}$ of a black hole horizon in the
evolution of STPv2 initial data, minus the Schwarzschild value of 0.5.
On the left are values at $t/M=0$.
On the right are values at $t_\text{relaxed}/M=250$.
}
\end{figure*}

Each initial data set is evolved on four different resolutions,
N0, N1, N2, and N3. These correspond to approximately $65^3$,
$71^3$, $78^3$, and $85^3$ grid points, respectively.
%The distribution of these grid points have not been optimized in
%any special way.
During the evolutions, neither the Hamiltonian and momentum constraints, 
nor the secondary constraints of the first-order generalized 
harmonic evolution system are explicitly enforced. 
By monitoring the constraint violations, we can obtain an 
indication of the accuracy of the evolutions.
Their values are shown in Fig.~\ref{fig:ConstraintsEv}.
Plotted is the $L^2$ norm of all the constraint fields of the 
first-order generalized harmonic system, normalized by the $L^2$ 
norm of the spatial gradients of the dynamical fields 
(cf. Eq.~(71) in~\cite{Lindblom2006}).  
The $L^2$ norms are taken over the portion of the computational volume
that lies outside the apparent horizons.
Below, we present the results from our evolutions with the 
highest resolution N3.

%\begin{figure}[h]
%\centering
%\begin{subfigure}
%  \centering
%  \includegraphics[width=10cm, trim=10cm 3cm 2.5cm 0cm]{HorizonCFMSMinusHalf_t0}
%  \caption{A subfigure}
%  \label{fig:sub1}
%\end{subfigure}%
%\begin{subfigure}
%  \centering
%  \includegraphics[width=10cm, trim=10cm 3cm 2.5cm 0cm]{HorizonCFMSMinusHalf_t250}
%  \caption{A subfigure}
%  \label{fig:sub2}
%\end{subfigure}
%\caption{A figure with subfigures}
%\label{fig:test}
%\end{figure}

%-------------------------------------------
\subsection{Horizon properties}
\label{sec:HorizonProps}
%-------------------------------------------
One would like to capture the correct horizon geometries as much as 
possible in the initial data, in order to avoid transients at early times
as the black holes relax.
We can visualize the deformations that a black hole undergoes by plotting 
the intrinsic scalar curvature $\bar{R}$ of its apparent horizon at 
various times.
This quantity is computed as
\begin{equation}
\bar{R} = R-2R_{ij}s^i s^j-\bar{K}^2+\bar{K}^{ij}\bar{K}_{ij},
\end{equation}
where
\begin{equation}
\bar{K}_{ij}=\nabla_is_j-s_is^k\nabla_ks_j
\end{equation}
is the extrinsic curvature of the apparent horizon embedded in $\Sigma_t$, 
and $s^i$ is the spatial unit normal to the apparent horizon.

\begin{figure}[t]
\includegraphics[scale=0.5]{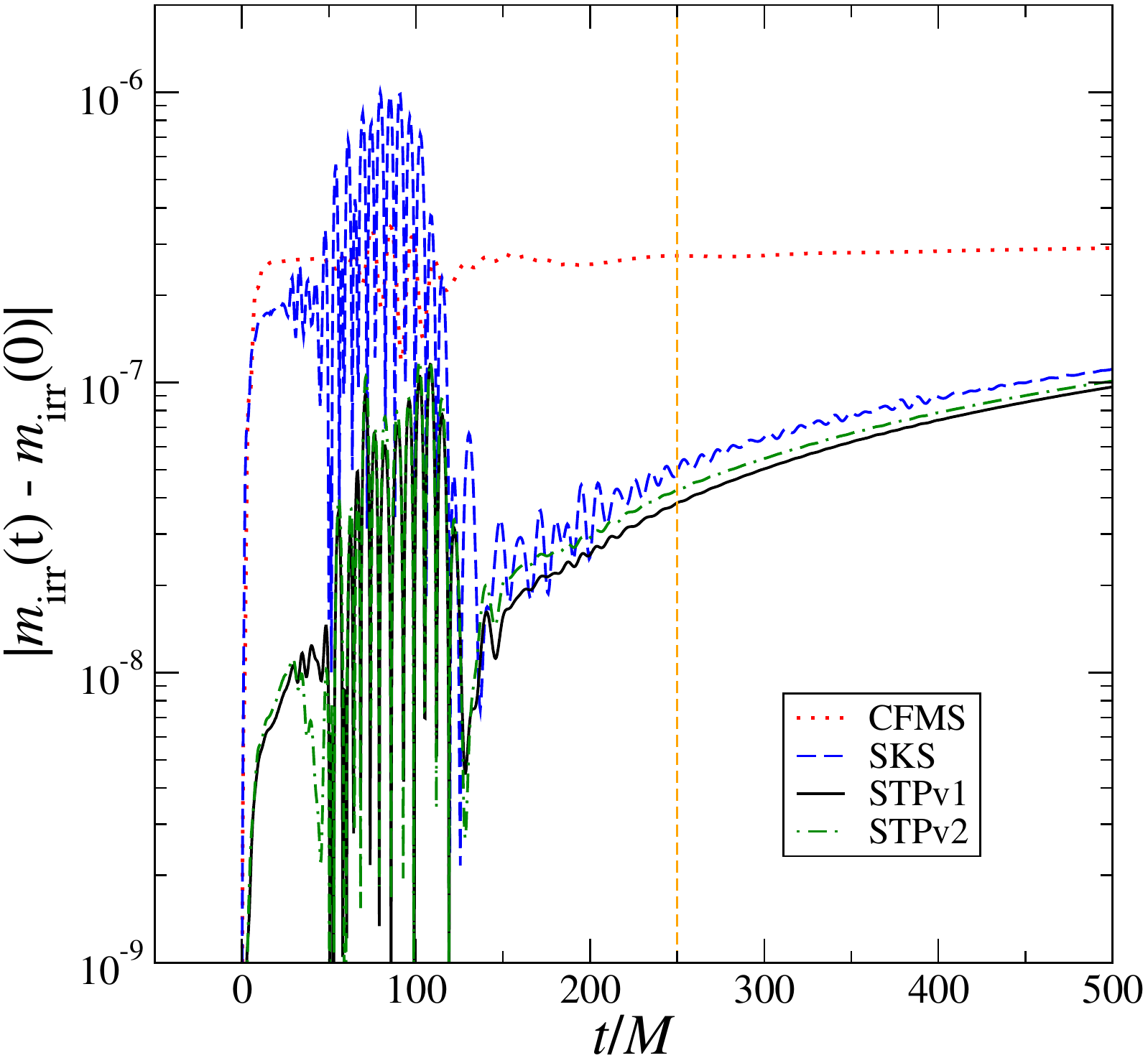}
\caption{
\label{fig:InitialMassDiffs}
Changes in the irreducible mass $m_\text{irr}(t)$ of a black hole
relative to its starting value, during the evolution of different initial
data sets. The vertical dashed line indicates a
time $t_\text{relaxed}/M=250$ after relaxation.
}
\end{figure}

Even though our STP initial data do include realistic tidal deformations, 
they do so only approximately. 
On the other hand, CFMS and SKS initial data do not take into account 
the tidal deformations at all, so it is expected that their black hole  
geometries will be farther from equilibrium. 
This is confirmed in Fig.~\ref{fig:HorizonCurvatureComparison}, which 
shows the extrema of $m^2\bar{R}$ over the surface of one black hole 
in the evolution of each initial data set. 
The horizon curvatures in the evolutions of CFMS and SKS initial data 
undergo similarly large variations at early times, but those for STP 
initial data are markedly smaller by about an order of magnitude.
All the extrema are centered around the value of 0.5, that for 
a single Schwarzschild black hole.

The values of $m^2\bar{R}$, minus the Schwarzschild value of 0.5,  
are plotted over the surfaces of the apparent horizons in 
Figs.~\ref{fig:HorizonComparisonCFMS} to~\ref{fig:HorizonComparisonSTPv2}
for a black hole in each initial data set,
and for the same black hole at a time after relaxation, which we take
to be $t_\text{relaxed}/M=250$.
(The other black hole, which is not shown, is on the negative $x$-axis.)
It is obvious that the black hole geometries in CFMS and SKS initial data
do not represent their relaxed values.
In contrast, the correct tidal deformations are much better captured in
STP initial data.

\begin{figure}[t]
\includegraphics[scale=0.5]{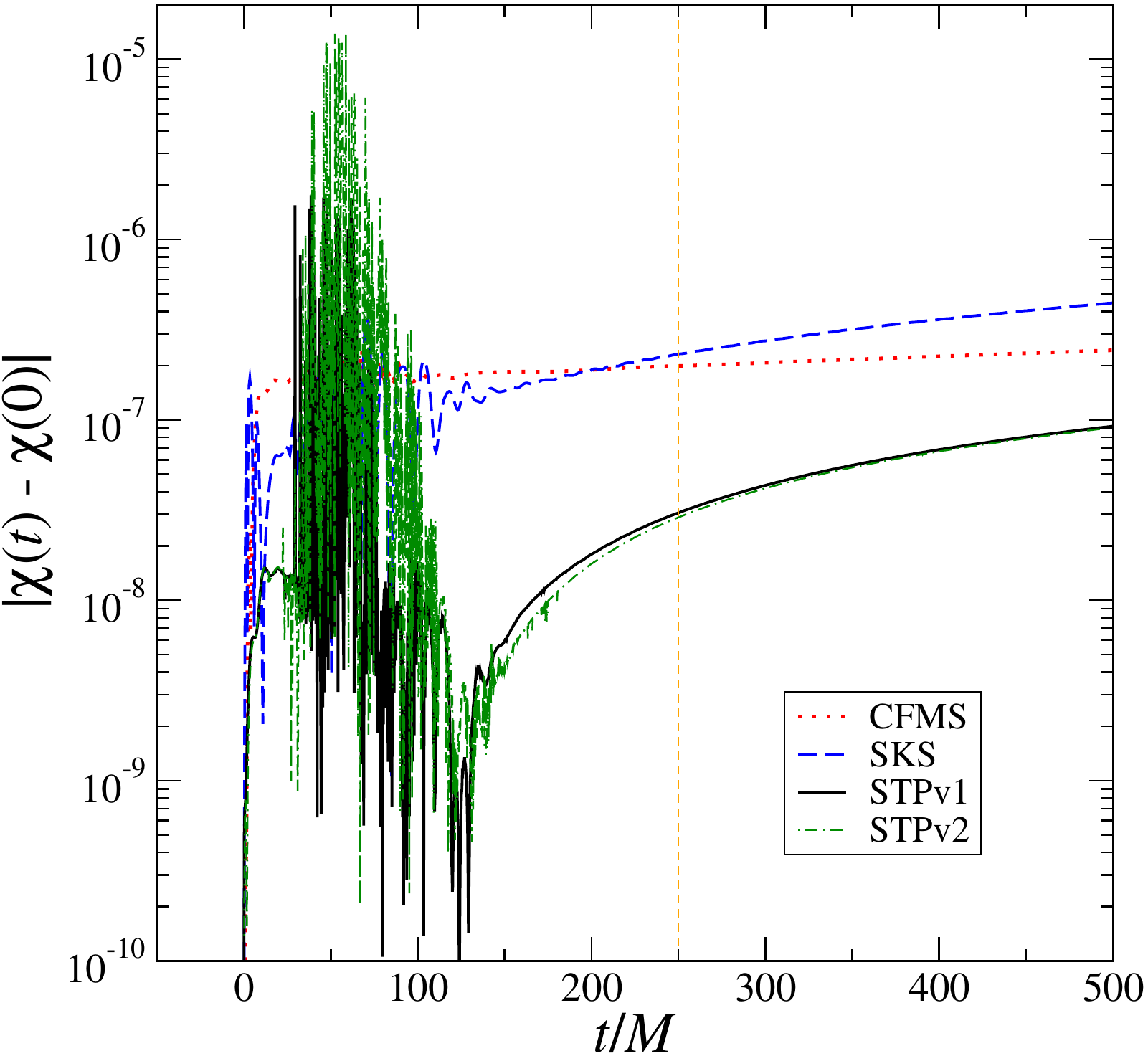}
\caption{
\label{fig:InitialSpinDiffs}
Changes in the dimensionless spin $\chi(t)$ of a black hole
relative to its starting value, during the evolution of different initial
data sets. The vertical dashed line indicates a
time $t_\text{relaxed}/M=250$ after relaxation.
}
\end{figure}

Changes in the black holes' masses and spins are also induced by 
the initial relaxation.
In Fig.~\ref{fig:InitialMassDiffs}, we show the difference in the 
irreducible mass from its initial value for a black hole in each evolution.
The irreducible mass is defined as~\cite{Hawking1968}
\begin{equation}
\label{eq:IrrMass}
m_\text{irr}(t) = \sqrt{A/16\pi},
\end{equation}
where $A$ is the area of the apparent horizon.
Near the time of relaxation $t_\text{relaxed}/M=250$, we see that 
going beyond conformal flatness with SKS initial data yields a 
noticeably smaller change in $m_\text{irr}(t)$ from its starting value. 
By including more realistic tidal deformations in our initial data, 
we can reduce this change slightly relative to SKS initial data. 
This is already evident with the inclusion of the Newtonian 
electric quadrupole in STPv2 initial data. 
By adding the higher-order pieces in STPv1 initial data, we 
achieve another slight improvement. 

In Fig.~\ref{fig:InitialSpinDiffs}, we show the difference in the
dimensionless spin $\chi(t)$ of a black hole in each evolution, 
computed over the apparent horizon using
approximate Killing vectors~\cite{Lovelace2008}, from its
starting value. Unlike the mass, the change in $\chi(t)$ for 
SKS initial data is comparable to that for CFMS initial data at 
$t_\text{relaxed}/M=250$.
In contrast, STP initial data show a clear improvement over
the others. By adding pieces of the tidal fields beyond the 
Newtonian electric quadrupole though, the change in $\chi(t)$ for 
STPv1 initial data is a bit larger than for STPv2 initial data.
This could be due to the corotating time dependence of the tidal 
fields that was used in STPv2 initial data, whereas STPv1 
initial data only included the linearized time dependence.

\begin{figure}[t]
\includegraphics[scale=0.5]{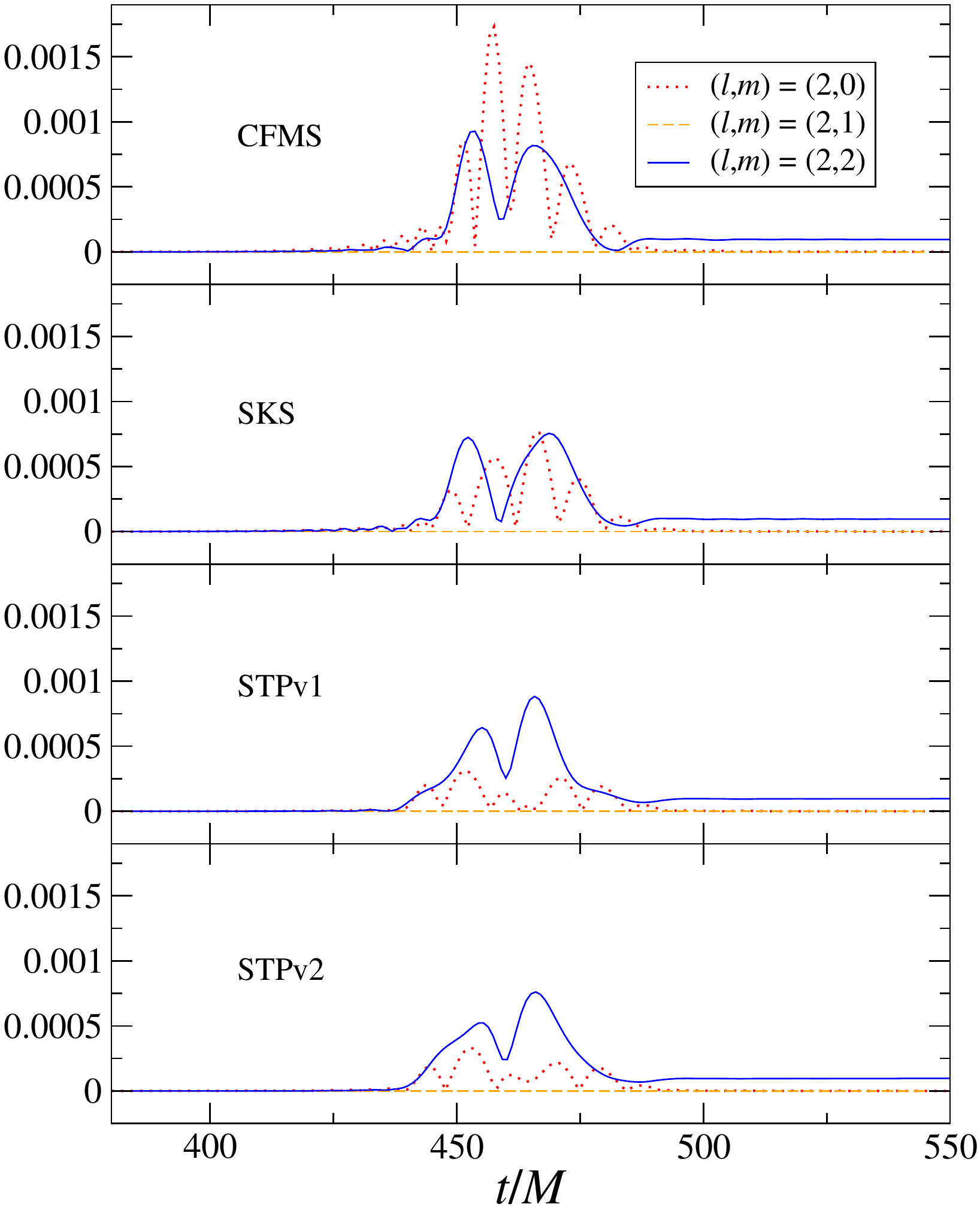}
\caption{
\label{fig:JunkPowerComparisonl2}
Junk radiation generated in the evolutions of different initial data sets,
shown for the $\left(l=2,m\geq0\right)$ modes of
$r_\text{extr}M\left|\Psi_4^{lm}\right|$ extracted at $r_\text{extr}/M=450$.
}
\end{figure}

It is interesting that while the rates of change in 
$m_\text{irr}(t)$ and $\chi(t)$ are similar after relaxation for 
SKS and STP initial data, the rate for CFMS initial data is more gradual. 
This may be caused by the particular nature of junk radiation arising from  
these types of initial data, such as where it predominantly originates and 
how much is ingoing or outgoing. 
A comparison of these rates with post-Newtonian predictions of tidal 
heating would also give further insight~\cite{Poisson2008}.
We quantify the amount of junk radiation 
generated in each initial data set in the next subsection, 
although determining their more detailed features requires further study.

\begin{figure}[t]
\includegraphics[scale=0.5]{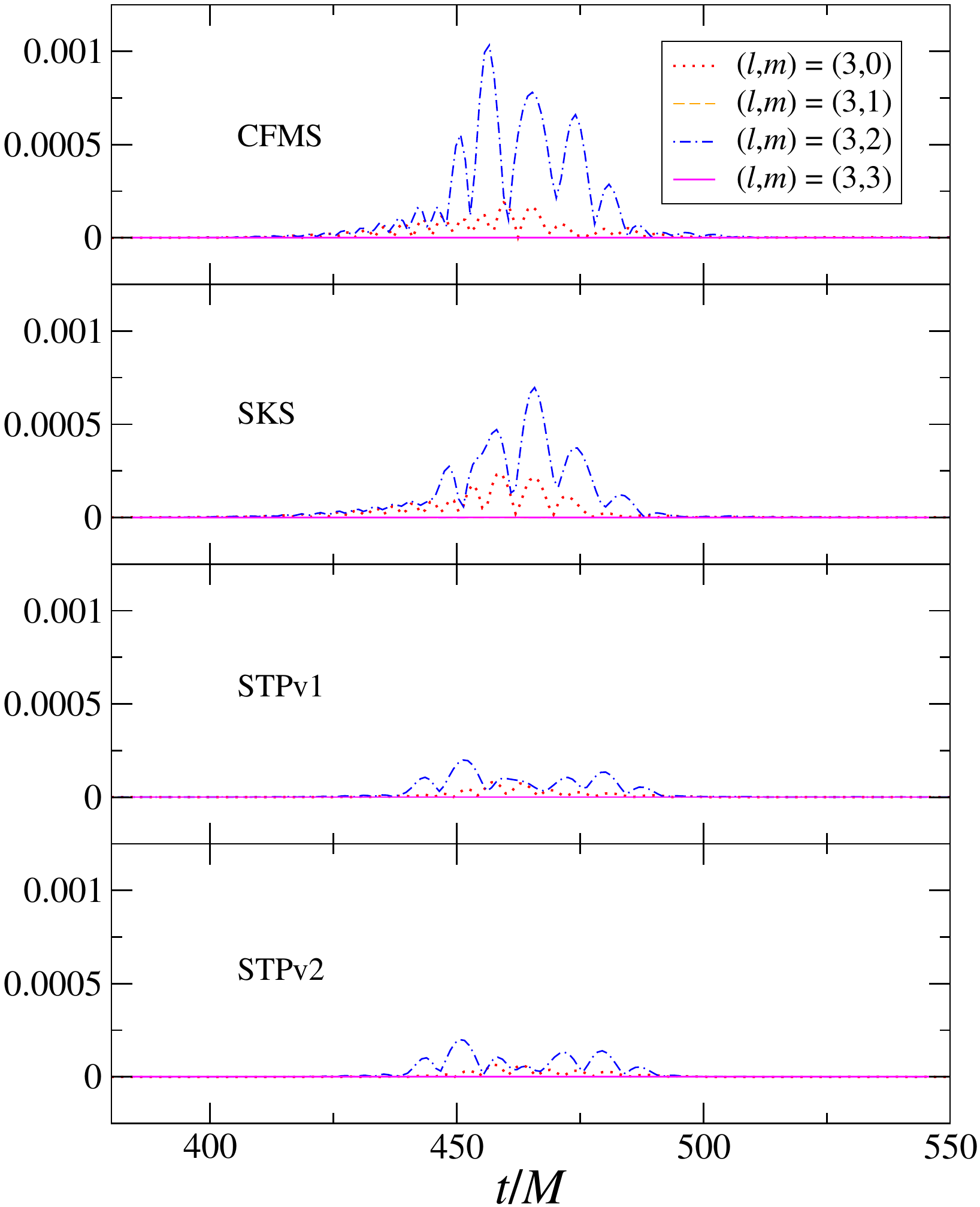}
\caption{
\label{fig:JunkPowerComparisonl3}
Junk radiation generated in the evolutions of different initial data sets,
shown for the $\left(l=3,m\geq0\right)$ modes of
$r_\text{extr}M\left|\Psi_4^{lm}\right|$ extracted at $r_\text{extr}/M=450$.
}
\end{figure}

%-------------------------------------------
\subsection{Junk radiation}
\label{sec:JunkRad}
%-------------------------------------------
At the beginning of an evolution, the entire spacetime geometry relaxes, 
and not just in the vicinity of the black holes. In the process, 
physically unrealistic gravitational radiation, or junk radiation, 
is generated. This junk radiation also contributes to the changes in 
the black holes discussed above. Here we compute and quantify the 
amount of junk radiation that develops for each initial data set.

Gravitational waves are extracted from the simulation on 
spheres of coordinate radius $r_{\text{extr}}$, 
following the same procedure in~\cite{Boyle2007}. 
We compute the Newman-Penrose scalar $\Psi_4$ given by 
\begin{equation}
\Psi_4=-C_{\alpha\mu\beta\nu}l^{\mu}l^{\nu}\bar{m}^{\alpha}\bar{m}^{\beta},
\end{equation} 
where $C_{\alpha\beta\gamma\delta}$ is the Weyl curvature tensor, and
\begin{align}
l^{\mu} &= \frac{1}{\sqrt{2}}\left(n^{\mu}-r^{\mu}\right), \\
\bar{m}^{\mu} &= \frac{1}{\sqrt{2}r}\left(\frac{\partial}{\partial\theta}-i\frac{1}{\sin\theta}\frac{\partial}{\partial\phi}\right)^{\mu},
\end{align}
in terms of spherical coordinates $\left(r,\theta,\phi\right)$ in the 
inertial frame, the timelike unit normal $n^{\mu}$ to the spatial 
hypersurface $\Sigma_t$, and the outward-pointing unit normal 
$r^{\mu}$ to the extraction sphere. 
Then $\Psi_4$ is expanded
in terms of spin-weighted spherical harmonics of weight $-2$,
\begin{equation}
\Psi_4\left(t,r,\theta,\phi\right)=\displaystyle\sum_{lm} 
\Psi_4^{lm}\left(t,r\right)_{-2}Y_{lm}\left(\theta,\phi\right),
\end{equation}
with expansion coefficients $\Psi_4^{lm}$.

The burst of junk radiation is visible at early times in $\Psi_4^{lm}$.
We consider the junk radiation extracted at $r_\text{extr}/M=450$. 
This is shown in Fig.~\ref{fig:JunkPowerComparisonl2} for the 
$\left(l=2,m\geq0\right)$ modes 
of $r_\text{extr}M\left|\Psi_4^{lm}\right|$.
For CFMS initial data, the largest component of the junk radiation is 
the $\left(2,0\right)$ mode. However, this mode is reduced by a factor of 
$\sim$2 each time, as we go to SKS and then to STP initial data.
For SKS initial data, the magnitudes of the $\left(2,2\right)$ and 
$\left(2,0\right)$ modes are comparable. For both STPv1 and 
STPv2 initial data, the magnitude of the $\left(2,0\right)$ mode is 
$\sim$2 times smaller than for the $\left(2,2\right)$ mode. 
There is no substantial reduction in the 
$\left(2,2\right)$ mode of the junk radiation for either SKS or STP 
initial data though. This is not surprising, since the $\left(2,2\right)$ 
mode has a lower frequency and is thought to be associated with the 
lack of outgoing gravitational radiation in the initial data, which 
is an issue present in all of our initial data sets.

\begin{figure}[t]
\includegraphics[scale=0.5]{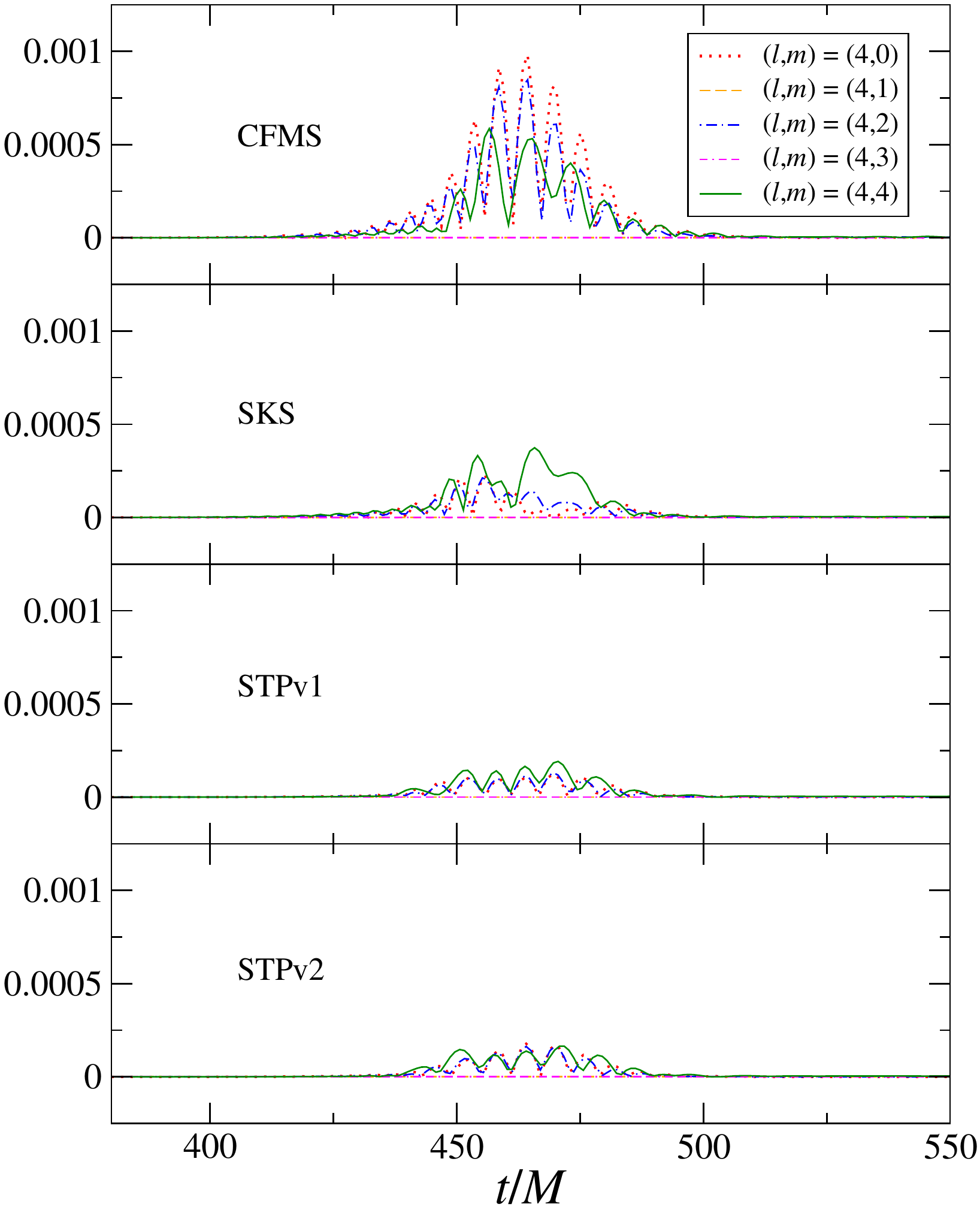}
\caption{
\label{fig:JunkPowerComparisonl4}
Junk radiation generated in the evolutions of different initial data sets,
shown for the $\left(l=4,m\geq0\right)$ modes of
$r_\text{extr}M\left|\Psi_4^{lm}\right|$ extracted at $r_\text{extr}/M=450$.
}
\end{figure}

Evidently, including realistic tidal deformations is more 
effective at reducing the 
higher frequency modes of the junk radiation caused by the initial 
ringing of the black holes. This is further illustrated for the 
$\left(l=3,m\geq0\right)$ modes in Fig.~\ref{fig:JunkPowerComparisonl3}.
There is a moderate reduction in the $\left(3,2\right)$ mode for SKS 
initial data over CFMS initial data, although there is  
no appreciable improvement for the $\left(3,0\right)$ mode. 
However, the $\left(3,2\right)$ mode for STP initial data is reduced by 
a factor of $\sim$4, and the $\left(3,0\right)$ mode by a factor of 
$\sim$2. A similar trend is seen for the $l=4$ modes in 
Fig.~\ref{fig:JunkPowerComparisonl4}, but in this case the decrease
in junk radiation for SKS initial data over CFMS initial data is much 
more obvious. Again, there are no major differences between STPv1 
and STPv2 initial data.

\begin{figure}[t]
\includegraphics[scale=0.5]{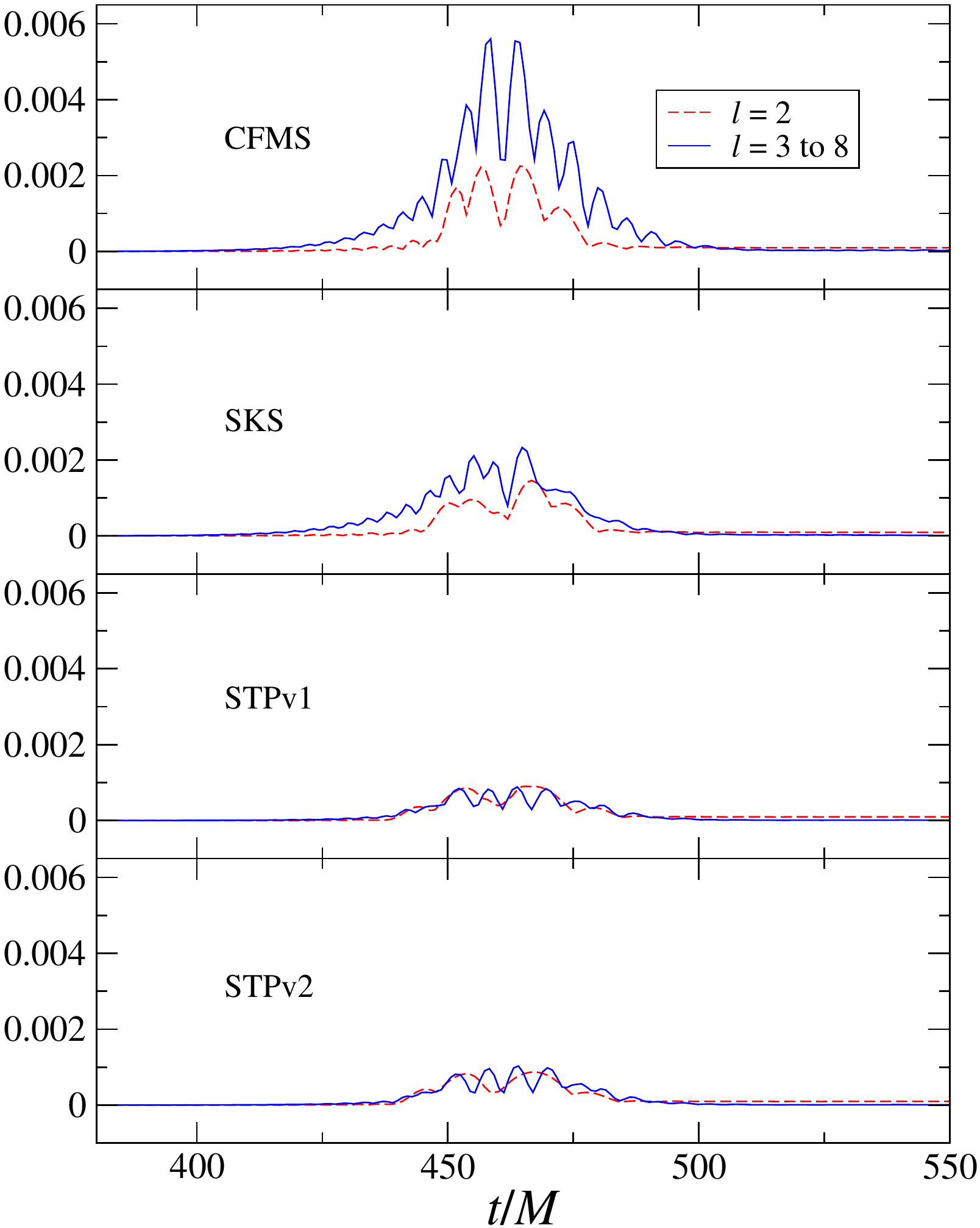}
\caption{
\label{fig:JunkPowerComparisonl2_l3+}
Total junk radiation generated in the evolutions of different initial data sets,
shown for the $\left(l=2,m\geq0\right)$ and
$\left(8\geq l\geq3,m\geq0\right)$ modes of
$r_\text{extr}M\left|\Psi_4^{lm}\right|$ extracted at $r_\text{extr}/M=450$.
%That is, we plot $\sum\limits_{l=3}^8\sum\limits_{m=0}^l%
%r_\text{extr}M\left|\Psi_4^{lm}\right|$ and
%$\sum\limits_{m=0}^2 r_\text{extr}M\left|\Psi_4^{2m}\right|$.
}
\end{figure}

To examine the other higher-order modes in the junk radiation, 
we show the total contribution in the $\left(8\geq l\geq3,m\geq0\right)$ 
modes in Fig.~\ref{fig:JunkPowerComparisonl2_l3+}.
We also plot the total contribution in only the 
$\left(l=2,m\geq0\right)$ modes for comparison. 
For both CFMS and SKS initial data, the combined junk radiation content 
in the higher-order modes is actually greater than in the $l=2$ modes alone, 
indicating the relevance of taking them into consideration. 
With our STP initial data, the situation is ameliorated, with the 
higher-order modes of the junk radiation being about the same 
magnitude as the $l=2$ modes. Simultaneously, both portions of the 
junk radiation are less than for CFMS and SKS initial data.

As a measure of the cumulative junk radiation content, we integrate 
the quantities in Fig.~\ref{fig:JunkPowerComparisonl2_l3+} over the time 
interval shown, $380M\leq t\leq 550M$, to obtain
\begin{align}
\label{eq:l2_JunkEnergy}
I_2 &:= \int_{380M}^{550M} \sum\limits_{m=0}^2 %
r_\text{extr}M\left|\Psi_4^{2m}\right| dt,\\
\label{eq:l3+_JunkEnergy}
I_{3+} &:= \int_{380M}^{550M} \sum\limits_{l=3}^8\sum\limits_{m=0}^l%
r_\text{extr}M\left|\Psi_4^{lm}\right| dt.
\end{align}
(An alternative would be to calculate the energy in the junk radiation, 
but we use this simpler measure here.)
Their values are displayed in Table~\ref{tab:JunkEnergy}.
The error estimates in $I_2$ and $I_{3+}$  are computed from the differences 
between the highest two resolutions, N2 and N3. For SKS initial data, 
$I_2$ is less by a factor of $\sim$1.4 and $I_{3+}$ is less by a factor 
of $\sim$2, relative to CFMS initial data. 
For both versions of STP initial data, $I_2$ is less by a factor of 
$\sim$1.7 and $I_{3+}$ is less by a factor of $\sim$5. 
They do differ somewhat from each other though, with $I_2$ a bit larger, 
and $I_{3+}$ a bit smaller, for STPv1 initial data. 
This again may be related to the different time dependences used in 
the tidal fields for the two cases.

\begin{table}[t]
\begin{tabular}{| >{\centering\arraybackslash}m{2cm} | >{\centering\arraybackslash}m{3cm} | >{\centering\arraybackslash}m{3cm} >{\centering\arraybackslash}m{0.00000001cm}|}
\hline
{\bf Initial data} & $\boldsymbol{I_2}$ & $\boldsymbol{I_{3+}}$ & \\[0.1cm]
\hline
CFMS & $0.050\pm0.002$ & $0.129\pm 0.004$ & \\[0.1cm]
SKS  & $0.0359\pm 0.0001$ & $0.064\pm 0.002$ & \\[0.1cm]
STPv1  & $0.0294\pm 0.0002$ & $0.0251\pm 0.0002$ & \\[0.1cm]
STPv2  & $0.0286\pm 0.0001$ & $0.0270\pm 0.0001$ & \\[0.1cm]
\hline
\end{tabular}
\caption{
Cumulative junk radiation content in the $\left(l=2,m\geq0\right)$ modes 
$I_2$, and in the $\left(8\geq l\geq3,m\geq0\right)$ modes $I_{3+}$, 
computed via Eqs.~\eqref{eq:l2_JunkEnergy}--\eqref{eq:l3+_JunkEnergy}.
\label{tab:JunkEnergy}}
\end{table}

%##############################################################################
\section{Discussion}
\label{sec:Discussion}
We have made a first attempt to include realistic tidal deformations in 
constraint-satisfying binary black-hole initial data, with the goal of 
understanding their effects on the relaxation of horizon properties, and 
the development of junk radiation in subsequent evolutions. 
This was done by superposing tidally perturbed 
black hole metrics as the conformal metric, in which the tidal fields were 
determined by asymptotically matching to a PN near zone 
metric~\cite{JohnsonMcDaniel:2009dq}. The results from evolutions were 
contrasted with those obtained with the widespread choice of conformally flat 
initial data, and those with initial data constructed by superposing 
Kerr-Schild metrics.  

By more accurately representing the horizon geometries in our initial data, 
we found that the black holes' intrinsic scalar curvatures 
deviated much less from their starting values at early times, with a 
corresponding decrease in the changes in the masses and spins.
Past studies~\cite{Kelly2010,Reifenberger2012}, 
also considering equal-mass and nonspinning black 
holes, have focused on the impact of including PN corrections 
on the dominant $\left(2,2\right)$ mode of the relaxed binary.
Here we have demonstrated that the total amount of junk radiation in the 
higher-order modes actually exceeds that in the $l=2$ modes alone, 
if one neglects realistic tidal deformations. 
With our STP initial data though, the junk radiation content 
in the higher-order modes becomes less than that in the $l=2$ modes, 
which is itself lowered. Thus, a more careful treatment of the horizons is 
not an inconsequential part of reducing junk radiation.

We constructed two versions of initial data with realistic tidal 
deformations, STPv1 that included all the 
tidal fields of Eqs.~\eqref{eq:TidalFields}--\eqref{eq:TidalFieldsLast}, 
and STPv2 that included only the Newtonian electric quadrupole but with 
a corotating time dependence given by 
Eqs.~\eqref{eq:xcorotating}--\eqref{eq:ycorotating}.
Both of these versions gave very similar results for the early 
relaxation of the black holes, and the amount of junk radiation generated. 
This suggests that incorporating only the lowest-order tidal fields 
may be an adequate treatment of the horizons, in the absence of further 
refinements to other parts of the spacetime geometry. 
Moreover, this lends support to the effectiveness of using the tidally 
perturbed spinning metrics of~\cite{Gallouin:2012}, 
which only have the lowest-order tidal fields. 
In Table~\ref{tab:JunkEnergy}, we also see that the cumulative 
junk radiation content in the $l=2$ modes is slightly larger for 
STPv1 initial data, but slightly smaller in the higher-order modes.
Perhaps this indicates that our STPv1 initial data will benefit by using 
a corotating time dependence of the tidal fields instead.

There are still many potential avenues to improve upon our initial data.
For instance, the superposition method itself limits how 
closely we are able to model 
the tidal deformations, because the Gaussian functions artificially 
alter the black hole metrics, and lead to unwanted perturbations 
of the horizon geometries. To overcome this, more sophisticated 
functions could be experimented with in the superposition, such as 
transition functions that satisfy the so-called Frankenstein 
theorems~\cite{Yunes2007}. 

Two of the most obvious shortcomings of our current initial data 
are the lack of outgoing gravitational radiation from the 
past history of the binary, and the absence of any interaction 
terms in the conformal metric for the region between the black holes. 
In fact, these were already accounted for in the approximate 
initial data of~\cite{JohnsonMcDaniel:2009dq}, which can be directly 
supplied as conformal data to solve for the Einstein constraint 
equations. This work is currently in progress~\cite{ThroweInPrep}.

It is straightforward to generalize our tidally perturbed initial data 
to unequal-mass binaries. 
Since the PN approximation is known to be less accurate in this 
regime~\cite{MacDonald:2012mp}, assuming the same binary separation, 
it would be interesting to see how well the results compare to the ones 
presented here. The effects of spin can also be investigated with the 
black hole metrics of~\cite{Gallouin:2012}. As mentioned earlier, we 
may also be able to reduce the eccentricity by adjusting the free parameters 
in the superposition functions, the metrics in Cook-Scheel coordinates, 
and the accompanying coordinate transformation to PN harmonic coordinates.  
Our initial data can then be combined with the iterative prescription 
of~\cite{Zhang:2013gda} to jointly diminish junk radiation and 
eccentricity. Finally, since our waveforms were extracted at a finite 
radius, they inevitably contain gauge effects. We have compared the 
junk radiation content for waveforms extracted at different radii, and find 
no significant differences. Nevertheless, it might be worthwhile 
to analyze the junk radiation with these effects removed through 
Cauchy-characteristic extraction, for 
instance~\cite{Winicour2009,Taylor:2013zia}.

%##############################################################################

\acknowledgements{
  We thank Keith D. Matthews, Harald P. Pfeiffer, 
  Mark A. Scheel, and B\'ela Szil\'agyi for helpful discussions. 
  %and to B\'ela Szil\'agyi for providing the spectral adaptive 
  %mesh refinement code for the evolutions with {\tt SpEC}.
  We especially appreciate Nathan K. Johnson-McDaniel's 
  many insights into the initial data of~\cite{JohnsonMcDaniel:2009dq}, 
  and his feedback on an earlier version of this manuscript. 
  We gratefully acknowledge support from the Natural Sciences and 
  Engineering Research Council (NSERC) of Canada, from the 
  Canada Research Chairs Program, and from the Canadian Institute for 
  Advanced Research.
  Initial data were constructed on the Zwicky cluster at Caltech, 
  which is supported by the Sherman Fairchild Foundation and  
  NSF award No. PHY-0960291.
  Evolutions were performed on the supercomputer 
  Briar\'ee at the University of Montreal, 
  under the administration of Calcul Qu\'ebec and Compute Canada, 
  and funded by the Canadian Foundation for Innovation (CFI), 
  NanoQu\'ebec, RMGA, and the Fonds de recherche du Qu\'ebec - 
  Nature et technologies (FRQ-NT).
  }  

\appendix*
\section{Tidally perturbed black hole metrics from asymptotic matching}

The exact expressions from~\cite{JohnsonMcDaniel:2009dq} that we use 
to construct our superposed tidally-perturbed initial data are collected 
here for completeness.
For all equations in this appendix, indices are raised and lowered with 
the flat spacetime metric $\eta_{\mu\nu}$.
One begins with a perturbed Schwarzschild metric in Cook-Scheel harmonic 
coordinates $X^\mu=\left(T,X^i\right)$~\cite{cook_scheel97}, 
comoving with and centered on that black hole,
\begin{widetext}
\begin{align}
\label{eq:CSmetric}
h_{\mu\nu}dX^{\mu}dX^{\nu} &= -H_{T^2}dT^2+H_{RT}dRdT
  +\frac{16}{3}\frac{\tilde{M}^2}{R}\left[1+\frac{\tilde{M}}{R}-\frac{2}{3}\frac{\tilde{M}^3}{R^2(R+\tilde{M})}\right]\dot{{\mathcal C}}_{klp}X^lX^pdX^kdT \notag\\
  &\quad+H_k^{[1]}dX^k\left[\left(1-\frac{\tilde{M}^2}{R^2}\right)dT-4\frac{\tilde{M}^2}{R^2}dR\right]+H_k^{[2]}dX^kdR+H_{R^2}dR^2+H_{\text{trc}}dX_sdX^s, \notag\\
  &\quad+{\mathcal O}(R^4/{\mathcal R}^4),
\end{align}
where $R:=\sqrt{X_iX^i}$ and the metric functions are
\begin{align}
\label{eq:CSmetricfunctions}
H_{T^2} &= \frac{R-\tilde{M}}{R+\tilde{M}}+\left[1-\frac{\tilde{M}}{R}\right]^2\left[({\mathcal E}_{kl}+T\dot{{\mathcal E}}_{kl})X^kX^l+\frac{1}{3}{\mathcal E}_{klp}X^kX^lX^p\right] \notag\\
  &\quad+\frac{4\tilde{M}^2}{(R+\tilde{M})^2}\left[R-\frac{5}{3}\frac{\tilde{M}^2}{R}\right]\dot{{\mathcal E}}_{kl}X^kX^l, \notag\\
H_{RT} &= \frac{8\tilde{M}^2}{(R+\tilde{M})^2}+8\frac{\tilde{M}^2}{R^2}\frac{R-\tilde{M}}{R+\tilde{M}}\left[({\mathcal E}_{kl}+T\dot{{\mathcal E}}_{kl})X^kX^l+\frac{1}{3}{\mathcal E}_{klp}X^kX^lX^p\right] \notag\\
  &\quad-\left[\frac{4}{3}R+\frac{14}{3}\tilde{M}+\frac{8}{3}\frac{\tilde{M}^2}{R}-2\frac{\tilde{M}^3}{R^2}-\frac{104}{3}\frac{\tilde{M}^4}{R^2(R+\tilde{M})}+\frac{80}{3}\frac{\tilde{M}^5}{R^2(R+\tilde{M})^2}\right. \notag\\
&\left. \quad\quad+\frac{32}{3}\frac{\tilde{M}^6}{R^2(R+\tilde{M})^3}\right]\dot{{\mathcal E}}_{kl}X^kX^l, \notag\\
H_k^{[1]} &= \frac{2}{3}\left[1+\frac{\tilde{M}}{R}\right]\left[2({\mathcal C}_{klp}+T\dot{{\mathcal C}}_{klp})X^lX^p+\left(1-\frac{1}{3}\frac{\tilde{M}}{R}\right){\mathcal C}_{klps}X^lX^pX^s\right], \\
H_k^{[2]} &= \left[\frac{R}{3}+2\tilde{M}+\frac{16}{3}\frac{\tilde{M}^2}{R}+\frac{26}{3}\frac{\tilde{M}^3}{R^2}-11\frac{\tilde{M}^4}{R^3}-\frac{32}{3}\frac{\tilde{M}^5}{R^3(R+\tilde{M})}-\frac{64}{9}\frac{\tilde{M}^6}{R^3(R+\tilde{M})^2}\right]\dot{{\mathcal C}}_{klp}X^lX^p, \notag\\
H_{R^2} &= \sum_{n=1}^3\left(\frac{2\tilde{M}}{R+\tilde{M}}\right)^n-\frac{2\tilde{M}}{R}-\frac{\tilde{M}^2}{R^2}+\left[2\frac{\tilde{M}}{R}+3\frac{\tilde{M}^2}{R^2}-\frac{\tilde{M}^4}{R^4}-\frac{16\tilde{M}^4}{R^2(R+\tilde{M})^2}\right]({\mathcal E}_{kl}+T\dot{{\mathcal E}}_{kl})X^kX^l, \notag\\
&\quad+\left[\frac{1}{3}\frac{\tilde{M}}{R}+\frac{1}{3}\frac{\tilde{M}^2}{R^2}-\frac{2}{5}\frac{\tilde{M}^3}{R^3}-\frac{7}{15}\frac{\tilde{M}^4}{R^4}-\frac{1}{15}\frac{\tilde{M}^5}{R^5}-\frac{16}{3}\frac{\tilde{M}^4}{R^2(R+\tilde{M})^2}\right]{\mathcal E}_{klp}X^kX^lX^p \notag\\
&\quad+\left[\frac{16}{3}\frac{\tilde{M}^2}{R}+\frac{80}{3}\frac{\tilde{M}^3}{R^2}+28\frac{\tilde{M}^4}{R^3}+\frac{40}{3}\frac{\tilde{M}^5}{R^4}-\frac{176}{3}\frac{\tilde{M}^6}{R^4(R+\tilde{M})}+\frac{72\tilde{M}^7}{R^4(R+\tilde{M})^2}\right. \notag\\
&\left.\quad\quad-\frac{32}{3}\frac{\tilde{M}^8}{R^4(R+\tilde{M})^3}-\frac{32}{3}\frac{\tilde{M}^9}{R^4(R+\tilde{M})^4}\right]\dot{{\mathcal E}}_{kl}X^kX^l, \notag\\
H_{\text{trc}} &= \left[1+\frac{\tilde{M}}{R}\right]^2\left[1-\left(1+2\frac{\tilde{M}}{R}-\frac{\tilde{M}^2}{R^2}\right)({\mathcal E}_{kl}+T\dot{{\mathcal E}}_{kl})X^kX^l\right. \notag\\
&\left.\quad-\frac{1}{3}\left(1+\frac{\tilde{M}}{R}-\frac{\tilde{M}^2}{R^2}-\frac{1}{5}\frac{\tilde{M}^3}{R^3}\right){\mathcal E}_{klp}X^kX^lX^p-4\frac{\tilde{M}^2}{R^2}\left(R+2\tilde{M}-\frac{2}{3}\frac{\tilde{M}^2}{R+\tilde{M}}\right)\dot{{\mathcal E}}_{kl}X^kX^l\right], \notag
\end{align}
\end{widetext}
and $\mathcal{R}$ is the characteristic length scale of the perturbation 
(see below). 
The electric quadrupole and octupole tidal fields are denoted by 
${\mathcal E}_{kl}$ and ${\mathcal E}_{klp}$, respectively. 
The magnetic quadrupole and octupole tidal fields are similarly 
denote by ${\mathcal B}_{kl}$ and ${\mathcal B}_{klp}$, respectively. 
They enter Eq.~\eqref{eq:CSmetric} through 
${\mathcal C}_{klp}=\epsilon_{kls}{\mathcal B}^s_p$ and
${\mathcal C}_{klps}=\epsilon_{klu}{\mathcal B}^u_{ps}$, where
$\epsilon_{ijk}$ is the spatial Levi-Civita symbol. The overdots on
the tidal fields denote time derivatives.
Note that $h_{\mu\nu}$ is formally only applicable for small $R$,
since the metric functions in Eq.~\eqref{eq:CSmetricfunctions}
contain terms that diverge as $R\rightarrow\infty$.

Next consider an $\mathcal{O}\left(v^4\right)$ PN 
metric~\cite{Blanchet-Faye-Ponsot:1998} 
in barycentric harmonic coordinates $x^\mu=\left(t,x^i\right)$ and 
specialized to a circular orbit, for which at $t=0$ one black hole 
(``hole 1") of mass $\tilde{m}_1$ lies along the positive $x$-axis 
and the other black hole (``hole 2") of mass $\tilde{m}_2$ lies along 
the negative $x$-axis. 
By asymptotically matching the perturbed Schwarzschild and PN metrics, 
the tidal fields about hole 1 are given by
\begin{widetext}
\begin{align}
\label{eq:TidalFields}
{\mathcal E}_{kl}(t) &= \frac{\tilde{m}_2}{b^3}\left\{\left[1-\frac{1}{2}\frac{\tilde{m}_2}{b}\right][\delta_{kl}-3\hat{x}_k\hat{x}_l]+\frac{1}{2}\frac{\tilde{m}}{b}[4\hat{x}_k\hat{x}_l-5\hat{y}_k\hat{y}_l+\hat{z}_k\hat{z}_l]-6\sqrt{\frac{\tilde{m}}{b}}\frac{t}{b}\hat{x}_{(k}\hat{y}_{l)}\right\}, \\
{\mathcal B}_{kl}(t) &= \frac{\tilde{m}_2}{b^3}\sqrt{\frac{\tilde{m}_2}{b}}\left\{\left[-6\sqrt{\frac{\tilde{m}}{\tilde{m}_2}}+\frac{\tilde{m}_2}{b}\left\{5\left(\frac{\tilde{m}}{\tilde{m}_2}\right)^{3/2}+7\sqrt{\frac{\tilde{m}}{\tilde{m}_2}}-3\sqrt{\frac{\tilde{m}_2}{\tilde{m}}}\right\}\right]\hat{x}_{(k}\hat{z}_{l)}\right. \notag\\ 
&\left. \quad-6\frac{\tilde{m}}{\tilde{m}_2}\sqrt{\frac{\tilde{m}_2}{b}}\frac{t}{b}\hat{y}_{(k}\hat{z}_{l)}\right\}, \\
{\mathcal E}_{klp}(t) &= \frac{\tilde{m}_2}{b^4}\left\{\left[1-3\frac{\tilde{m}_2}{b}\right][15\hat{x}_{k}\hat{x}_{l}\hat{x}_{p}-9\delta_{(kl}\hat{x}_{p)}]-3\frac{\tilde{m}}{b}[\hat{x}_k\hat{x}_l\hat{x}_p-4\hat{y}_{(k}\hat{y}_{l}\hat{x}_{p)}+\hat{z}_{(k}\hat{z}_l\hat{x}_{p)}]\right\}, \\
\label{eq:TidalFieldsLast}
{\mathcal B}_{klp}(t) &= \frac{9}{2}\frac{\tilde{m}_2}{b^4}\sqrt{\frac{\tilde{m}}{b}}\left[5\hat{x}_{(k}\hat{x}_l\hat{z}_{p)}-\delta_{(kl}\hat{z}_{p)}\right],
\end{align}
\end{widetext}
where $\tilde{m}=\tilde{m}_1+\tilde{m}_2$, $b$ is the PN coordinate separation 
of the holes, and $\hat{x}^{\mu},\hat{y}^{\mu}$, $\hat{z}^{\mu}$
(and $\hat{t}^{\mu}$ below, with $\hat{t}_0=-1$) are Cartesian basis vectors.
In the region around hole 1, the characteristic
length scale of the perturbation in Eq.~\eqref{eq:CSmetric} is
${\mathcal R}\sim\sqrt{b^3/m_2}$~\cite{Thorne1985}.
The tidal fields about hole 2 are obtained by letting 
$\tilde{m}_2\rightarrow \tilde{m}_1$,
$\hat{x}_{\mu}\rightarrow -\hat{x}_{\mu}$, and
$\hat{y}_{\mu}\rightarrow -\hat{y}_{\mu}$.
These tidal fields are only valid for times $t\approx0$. 

Finally, the perturbed Schwarzschild metrics are transformed to PN 
harmonic coordinates (only up to $\mathcal{O}\left(v^3\right)$ here), 
which around hole 1 is given by 
\begin{equation}
\label{eq:PNtoCS}
X^{\alpha}(x^{\beta}) = \sum_{j=0}^3 \left(\frac{\tilde{m}_2}{b}\right)^{j/2}\left(X^{\alpha}\right)_j(x^{\beta})+{\mathcal O}\left(v^4\right),
\end{equation}
where
\begin{align}
\left(X_{\alpha}\right)_0 &= x_{\alpha}+\left(\tilde{m}_2/\tilde{m}\right)b\hat{x}_{\alpha}, \\
\left(X_{\alpha}\right)_1 &= \left(F_{\beta\alpha}\right)_1\tilde{x}^{\beta}, \\
\left(X_{\alpha}\right)_2 &= \left[1-\frac{\tilde{x}}{b}\right]\Delta_{\alpha\beta}\tilde{x}^{\beta}+\frac{\Delta_{\beta\gamma}\tilde{x}^{\beta}\tilde{x}^{\gamma}}{2b}\hat{x}_{\alpha} \notag\\ 
\quad&-\frac{1}{2}\left(F_{\alpha}^{\;\gamma}\right)_1\left(F_{\beta\gamma}\right)_1\tilde{x}^{\beta}, \\
\left(X_{\alpha}\right)_3 &= \sqrt{\frac{\tilde{m}}{\tilde{m}_2}}\left\{-\frac{yt}{b^2}\Delta_{\alpha\beta}\tilde{x}^{\beta}+\left[\frac{\tilde{x}_{\mu}\tilde{x}^{\mu}-4\tilde{x}^2}{2b^2} \right.\right. \notag\\
  &\quad\left.+\left(2-\frac{\tilde{m}_2}{\tilde{m}}\right)\frac{\tilde{x}}{b}+\left(2+\frac{1}{2}\frac{\tilde{m}_2}{\tilde{m}}\right)\frac{\tilde{m}_2}{\tilde{m}}\right]y\hat{t}_{\alpha} \notag\\
  &\quad\left.+2\left[1-\frac{\tilde{m}_2}{\tilde{m}}\right]\frac{yt}{b}\hat{x}_{\alpha}+\left[\frac{3\tilde{r}^2+t^2}{6b^2} \right.\right. \notag\\
  &\quad\left.\left.+\left(\frac{\tilde{m}_2}{\tilde{m}}-2\right)\frac{\tilde{x}}{b}+\frac{1}{2}\left(\frac{\tilde{m}_2}{\tilde{m}}\right)^2+4\right]t\hat{y}_{\alpha}\right\} \notag \\
  &\quad+\left(F_{\beta\alpha}\right)_3\tilde{x}^{\beta}+\frac{1}{2b^3}\sqrt{\frac{\tilde{m}}{\tilde{m}_2}}\tilde{x}y\left(4\tilde{x}^2-y^2-z^2\right)\hat{t}_{\alpha}.
\end{align}
In the above,
$\tilde{x}^{\alpha}:=x^{\alpha}-\left(\tilde{m}_2/\tilde{m}\right)b\hat{x}^{\alpha}$, 
$\tilde{x}:=x-\left(\tilde{m}_2/\tilde{m}\right)b$, 
$\tilde{r}:=\sqrt{\tilde{x}_k\tilde{x}^k}$, 
$\Delta_{\alpha\beta}:=\text{diag}\left(1,1,1,1\right)$, and
\begin{align}
\left(F_{\alpha\beta}\right)_1 &= 2\sqrt{\frac{\tilde{m}_2}{\tilde{m}}}\hat{t}_{[\alpha}\hat{y}_{\beta]}, \\
\label{eq:PNtoCS_F3}
\left(F_{\alpha\beta}\right)_3 &= \left[\left(\frac{\tilde{m}_2}{\tilde{m}}\right)^{3/2}+3\sqrt{\frac{\tilde{m}_2}{\tilde{m}}}+5\sqrt{\frac{\tilde{m}}{\tilde{m}_2}}\right]\hat{t}_{[\alpha}\hat{y}_{\beta]},
\end{align}
encode the parts of the hole's Lorentz boost that are determined by the
matching.
The coordinate transformation around hole 2 is obtained from the preceding
expressions by making the substitutions 
$\tilde{m}_1\rightarrow \tilde{m}_2$,
$\left(t,x,y,z\right)\rightarrow\left(t,-x,-y,z\right)$, and
$\left(T,X,Y,Z\right)\rightarrow\left(T,-X,-Y,Z\right)$.

\vfill
%%%%%%%%%%%%%%%%%%%%%%%%%%%%%%%%%%%%%%%%%%%%%%%%%%%%%%%%%%%%%%%%%%%%%%%%%%%%%%%
%\section*{References}
%%%%%%%%%%%%%%%%%%%%%%%%%%%%%%%%%%%%%%%%%%%%%%%%%%%%%%%%%%%%%%%%%%%%%%%%%%%%%%%
\bibliography{References/References}

\begin{thebibliography}{71}
\expandafter\ifx\csname natexlab\endcsname\relax\def\natexlab#1{#1}\fi
\expandafter\ifx\csname bibnamefont\endcsname\relax
  \def\bibnamefont#1{#1}\fi
\expandafter\ifx\csname bibfnamefont\endcsname\relax
  \def\bibfnamefont#1{#1}\fi
\expandafter\ifx\csname citenamefont\endcsname\relax
  \def\citenamefont#1{#1}\fi
\expandafter\ifx\csname url\endcsname\relax
  \def\url#1{\texttt{#1}}\fi
\expandafter\ifx\csname urlprefix\endcsname\relax\def\urlprefix{URL }\fi
\providecommand{\bibinfo}[2]{#2}
\providecommand{\eprint}[2][]{\url{#2}}

\bibitem[{\citenamefont{et~al.}(2006)}]{ligoBBH}
\bibinfo{author}{\bibfnamefont{B.~A.} \bibnamefont{et~al.}}
  (\bibinfo{collaboration}{LIGO Scientific Collaboration}),
  \bibinfo{journal}{Phys.\ Rev.\ D} \textbf{\bibinfo{volume}{73}},
  \bibinfo{pages}{062001} (\bibinfo{year}{2006}).

\bibitem[{\citenamefont{di~Fiore}(2002)}]{Acernese:2002}
\bibinfo{author}{\bibfnamefont{L.}~\bibnamefont{di~Fiore}}
  (\bibinfo{collaboration}{{VIRGO}}), \bibinfo{journal}{Class.\ Quantum Grav.}
  \textbf{\bibinfo{volume}{19}}, \bibinfo{pages}{1421} (\bibinfo{year}{2002}).

\bibitem[{\citenamefont{Garat and Price}(2000)}]{GaratPrice:2000}
\bibinfo{author}{\bibfnamefont{A.}~\bibnamefont{Garat}} \bibnamefont{and}
  \bibinfo{author}{\bibfnamefont{R.~H.} \bibnamefont{Price}},
  \bibinfo{journal}{Phys.\ Rev.\ D} \textbf{\bibinfo{volume}{61}},
  \bibinfo{pages}{124011} (\bibinfo{year}{2000}).

\bibitem[{\citenamefont{{Valiente Kroon}}(2004)}]{Kroon:2004b}
\bibinfo{author}{\bibfnamefont{J.~A.} \bibnamefont{{Valiente Kroon}}},
  \bibinfo{journal}{Class.\ Quantum Grav.} \textbf{\bibinfo{volume}{21}},
  \bibinfo{pages}{3237} (\bibinfo{year}{2004}).

\bibitem[{\citenamefont{Nissanke}(2006)}]{Nissanke2006}
\bibinfo{author}{\bibfnamefont{S.}~\bibnamefont{Nissanke}},
  \bibinfo{journal}{Phys.\ Rev.\ D} \textbf{\bibinfo{volume}{73}},
  \bibinfo{pages}{124002} (\bibinfo{year}{2006}).

\bibitem[{\citenamefont{Boyle et~al.}(2007)\citenamefont{Boyle, Brown, Kidder,
  Mrou{\'e}, Pfeiffer, Scheel, Cook, and Teukolsky}}]{Boyle2007}
\bibinfo{author}{\bibfnamefont{M.}~\bibnamefont{Boyle}},
  \bibinfo{author}{\bibfnamefont{D.~A.} \bibnamefont{Brown}},
  \bibinfo{author}{\bibfnamefont{L.~E.} \bibnamefont{Kidder}},
  \bibinfo{author}{\bibfnamefont{A.~H.} \bibnamefont{Mrou{\'e}}},
  \bibinfo{author}{\bibfnamefont{H.~P.} \bibnamefont{Pfeiffer}},
  \bibinfo{author}{\bibfnamefont{M.~A.} \bibnamefont{Scheel}},
  \bibinfo{author}{\bibfnamefont{G.~B.} \bibnamefont{Cook}}, \bibnamefont{and}
  \bibinfo{author}{\bibfnamefont{S.~A.} \bibnamefont{Teukolsky}},
  \bibinfo{journal}{Phys.\ Rev.\ D} \textbf{\bibinfo{volume}{76}},
  \bibinfo{eid}{124038} (\bibinfo{year}{2007}).

\bibitem[{\citenamefont{Hannam et~al.}(2008)\citenamefont{Hannam, Husa,
  Gonz{\'a}lez, Sperhake, and Br{\"u}gmann}}]{Hannam2007}
\bibinfo{author}{\bibfnamefont{M.}~\bibnamefont{Hannam}},
  \bibinfo{author}{\bibfnamefont{S.}~\bibnamefont{Husa}},
  \bibinfo{author}{\bibfnamefont{J.~A.} \bibnamefont{Gonz{\'a}lez}},
  \bibinfo{author}{\bibfnamefont{U.}~\bibnamefont{Sperhake}}, \bibnamefont{and}
  \bibinfo{author}{\bibfnamefont{B.}~\bibnamefont{Br{\"u}gmann}},
  \bibinfo{journal}{Phys.\ Rev.\ D} \textbf{\bibinfo{volume}{77}},
  \bibinfo{pages}{044020} (\bibinfo{year}{2008}).

\bibitem[{\citenamefont{{Zlochower} et~al.}(2012)\citenamefont{{Zlochower},
  {Ponce}, and {Lousto}}}]{Zlochower2012}
\bibinfo{author}{\bibfnamefont{Y.}~\bibnamefont{{Zlochower}}},
  \bibinfo{author}{\bibfnamefont{M.}~\bibnamefont{{Ponce}}}, \bibnamefont{and}
  \bibinfo{author}{\bibfnamefont{C.~O.} \bibnamefont{{Lousto}}},
  \bibinfo{journal}{Phys.\ Rev.\ D} \textbf{\bibinfo{volume}{86}},
  \bibinfo{pages}{104056} (\bibinfo{year}{2012}).

\bibitem[{\citenamefont{Postnov and Yungelson}(2005)}]{Postnov:2007jv}
\bibinfo{author}{\bibfnamefont{K.}~\bibnamefont{Postnov}} \bibnamefont{and}
  \bibinfo{author}{\bibfnamefont{L.}~\bibnamefont{Yungelson}},
  \bibinfo{journal}{Living Rev. Rel.} \textbf{\bibinfo{volume}{9}},
  \bibinfo{pages}{6} (\bibinfo{year}{2005}), \eprint{astro-ph/0701059}.

\bibitem[{\citenamefont{Zhang and Szil\'agyi}(2013)}]{Zhang:2013gda}
\bibinfo{author}{\bibfnamefont{F.}~\bibnamefont{Zhang}} \bibnamefont{and}
  \bibinfo{author}{\bibfnamefont{B.}~\bibnamefont{Szil\'agyi}},
  \bibinfo{journal}{Phys.\ Rev.\ D} \textbf{\bibinfo{volume}{88}},
  \bibinfo{pages}{084033} (\bibinfo{year}{2013}).

\bibitem[{\citenamefont{Matzner et~al.}(1998)\citenamefont{Matzner, Huq, and
  Shoemaker}}]{Matzner1999}
\bibinfo{author}{\bibfnamefont{R.~A.} \bibnamefont{Matzner}},
  \bibinfo{author}{\bibfnamefont{M.~F.} \bibnamefont{Huq}}, \bibnamefont{and}
  \bibinfo{author}{\bibfnamefont{D.}~\bibnamefont{Shoemaker}},
  \bibinfo{journal}{Phys.\ Rev.\ D} \textbf{\bibinfo{volume}{59}},
  \bibinfo{pages}{024015} (\bibinfo{year}{1998}).

\bibitem[{\citenamefont{Marronetti and
  Matzner}(2000)}]{Marronetti-Matzner:2000}
\bibinfo{author}{\bibfnamefont{P.}~\bibnamefont{Marronetti}} \bibnamefont{and}
  \bibinfo{author}{\bibfnamefont{R.~A.} \bibnamefont{Matzner}},
  \bibinfo{journal}{Phys.\ Rev.\ Lett.} \textbf{\bibinfo{volume}{85}},
  \bibinfo{pages}{5500} (\bibinfo{year}{2000}).

\bibitem[{\citenamefont{Marronetti et~al.}(2000)\citenamefont{Marronetti, Huq,
  Laguna, Lehner, Matzner, and Shoemaker}}]{Marronetti2000}
\bibinfo{author}{\bibfnamefont{P.}~\bibnamefont{Marronetti}},
  \bibinfo{author}{\bibfnamefont{M.}~\bibnamefont{Huq}},
  \bibinfo{author}{\bibfnamefont{P.}~\bibnamefont{Laguna}},
  \bibinfo{author}{\bibfnamefont{L.}~\bibnamefont{Lehner}},
  \bibinfo{author}{\bibfnamefont{R.~A.} \bibnamefont{Matzner}},
  \bibnamefont{and}
  \bibinfo{author}{\bibfnamefont{D.}~\bibnamefont{Shoemaker}},
  \bibinfo{journal}{Phys.\ Rev.\ D} \textbf{\bibinfo{volume}{62}},
  \bibinfo{pages}{024017} (\bibinfo{year}{2000}).

\bibitem[{\citenamefont{Hannam et~al.}(2007)\citenamefont{Hannam, Husa,
  Br{\"u}gmann, Gonzalez, and Sperhake}}]{Hannam2007b}
\bibinfo{author}{\bibfnamefont{M.}~\bibnamefont{Hannam}},
  \bibinfo{author}{\bibfnamefont{S.}~\bibnamefont{Husa}},
  \bibinfo{author}{\bibfnamefont{B.}~\bibnamefont{Br{\"u}gmann}},
  \bibinfo{author}{\bibfnamefont{J.~A.} \bibnamefont{Gonzalez}},
  \bibnamefont{and} \bibinfo{author}{\bibfnamefont{U.}~\bibnamefont{Sperhake}},
  \bibinfo{journal}{Class.\ Quantum Grav.} \textbf{\bibinfo{volume}{24}},
  \bibinfo{pages}{S15} (\bibinfo{year}{2007}), \eprint{gr-qc/0612001}.

\bibitem[{\citenamefont{Lovelace et~al.}(2008)\citenamefont{Lovelace, Owen,
  Pfeiffer, and Chu}}]{Lovelace2008}
\bibinfo{author}{\bibfnamefont{G.}~\bibnamefont{Lovelace}},
  \bibinfo{author}{\bibfnamefont{R.}~\bibnamefont{Owen}},
  \bibinfo{author}{\bibfnamefont{H.~P.} \bibnamefont{Pfeiffer}},
  \bibnamefont{and} \bibinfo{author}{\bibfnamefont{T.}~\bibnamefont{Chu}},
  \bibinfo{journal}{Phys.\ Rev.\ D} \textbf{\bibinfo{volume}{78}},
  \bibinfo{pages}{084017} (\bibinfo{year}{2008}).

\bibitem[{\citenamefont{Lovelace}(2009)}]{Lovelace2009}
\bibinfo{author}{\bibfnamefont{G.}~\bibnamefont{Lovelace}},
  \bibinfo{journal}{Class.\ Quantum Grav.} \textbf{\bibinfo{volume}{26}},
  \bibinfo{pages}{114002} (\bibinfo{year}{2009}).

\bibitem[{\citenamefont{Blanchet}(2006)}]{Blanchet2006}
\bibinfo{author}{\bibfnamefont{L.}~\bibnamefont{Blanchet}},
  \bibinfo{journal}{Living Rev.Rel.} \textbf{\bibinfo{volume}{9}},
  \bibinfo{pages}{4} (\bibinfo{year}{2006}).

\bibitem[{\citenamefont{Kelly et~al.}(2007)\citenamefont{Kelly, Tichy,
  Campanelli, and Whiting}}]{kellyEtAl:2007}
\bibinfo{author}{\bibfnamefont{B.~J.} \bibnamefont{Kelly}},
  \bibinfo{author}{\bibfnamefont{W.}~\bibnamefont{Tichy}},
  \bibinfo{author}{\bibfnamefont{M.}~\bibnamefont{Campanelli}},
  \bibnamefont{and} \bibinfo{author}{\bibfnamefont{B.~F.}
  \bibnamefont{Whiting}}, \bibinfo{journal}{Phys.\ Rev.\ D}
  \textbf{\bibinfo{volume}{76}}, \bibinfo{pages}{024008}
  (\bibinfo{year}{2007}).

\bibitem[{\citenamefont{Kelly et~al.}(2010)\citenamefont{Kelly, Tichy,
  Zlochower, Campanelli, and Whiting}}]{Kelly2010}
\bibinfo{author}{\bibfnamefont{B.~J.} \bibnamefont{Kelly}},
  \bibinfo{author}{\bibfnamefont{W.}~\bibnamefont{Tichy}},
  \bibinfo{author}{\bibfnamefont{Y.}~\bibnamefont{Zlochower}},
  \bibinfo{author}{\bibfnamefont{M.}~\bibnamefont{Campanelli}},
  \bibnamefont{and} \bibinfo{author}{\bibfnamefont{B.}~\bibnamefont{Whiting}},
  \bibinfo{journal}{Class.\ Quantum Grav.} \textbf{\bibinfo{volume}{27}},
  \bibinfo{pages}{114005} (\bibinfo{year}{2010}).

\bibitem[{\citenamefont{Mundim et~al.}(2011)\citenamefont{Mundim, Kelly,
  Zlochower, Nakano, and Campanelli}}]{Mundim2011}
\bibinfo{author}{\bibfnamefont{B.~C.} \bibnamefont{Mundim}},
  \bibinfo{author}{\bibfnamefont{B.~J.} \bibnamefont{Kelly}},
  \bibinfo{author}{\bibfnamefont{Y.}~\bibnamefont{Zlochower}},
  \bibinfo{author}{\bibfnamefont{H.}~\bibnamefont{Nakano}}, \bibnamefont{and}
  \bibinfo{author}{\bibfnamefont{M.}~\bibnamefont{Campanelli}},
  \bibinfo{journal}{Class.\ Quantum Grav.} \textbf{\bibinfo{volume}{28}},
  \bibinfo{pages}{134003} (\bibinfo{year}{2011}).

\bibitem[{\citenamefont{Johnson-McDaniel
  et~al.}(2009)\citenamefont{Johnson-McDaniel, Yunes, Tichy, and
  Owen}}]{JohnsonMcDaniel:2009dq}
\bibinfo{author}{\bibfnamefont{N.~K.} \bibnamefont{Johnson-McDaniel}},
  \bibinfo{author}{\bibfnamefont{N.}~\bibnamefont{Yunes}},
  \bibinfo{author}{\bibfnamefont{W.}~\bibnamefont{Tichy}}, \bibnamefont{and}
  \bibinfo{author}{\bibfnamefont{B.~J.} \bibnamefont{Owen}},
  \bibinfo{journal}{Phys.Rev.} \textbf{\bibinfo{volume}{D80}},
  \bibinfo{pages}{124039} (\bibinfo{year}{2009}), \eprint{0907.0891}.

\bibitem[{\citenamefont{Gallouin et~al.}(2012)\citenamefont{Gallouin, Nakano,
  Yunes, and Campanelli}}]{Gallouin:2012}
\bibinfo{author}{\bibfnamefont{L.}~\bibnamefont{Gallouin}},
  \bibinfo{author}{\bibfnamefont{H.}~\bibnamefont{Nakano}},
  \bibinfo{author}{\bibfnamefont{N.}~\bibnamefont{Yunes}}, \bibnamefont{and}
  \bibinfo{author}{\bibfnamefont{M.}~\bibnamefont{Campanelli}},
  \bibinfo{journal}{Class.\ Quantum Grav.} \textbf{\bibinfo{volume}{29}},
  \bibinfo{pages}{235013} (\bibinfo{year}{2012}), \eprint{arXiv:1208.6489}.

\bibitem[{\citenamefont{Yunes et~al.}(2006)\citenamefont{Yunes, Tichy, Owen,
  and Br{\"u}gmann}}]{Yunes2006a}
\bibinfo{author}{\bibfnamefont{N.}~\bibnamefont{Yunes}},
  \bibinfo{author}{\bibfnamefont{W.}~\bibnamefont{Tichy}},
  \bibinfo{author}{\bibfnamefont{B.~J.} \bibnamefont{Owen}}, \bibnamefont{and}
  \bibinfo{author}{\bibfnamefont{B.}~\bibnamefont{Br{\"u}gmann}},
  \bibinfo{journal}{Phys.\ Rev.\ D} \textbf{\bibinfo{volume}{74}},
  \bibinfo{pages}{104011} (\bibinfo{year}{2006}).

\bibitem[{\citenamefont{Yunes and Tichy}(2006)}]{Yunes2006b}
\bibinfo{author}{\bibfnamefont{N.}~\bibnamefont{Yunes}} \bibnamefont{and}
  \bibinfo{author}{\bibfnamefont{W.}~\bibnamefont{Tichy}},
  \bibinfo{journal}{Phys.\ Rev.\ D} \textbf{\bibinfo{volume}{74}},
  \bibinfo{pages}{064013} (\bibinfo{year}{2006}).

\bibitem[{\citenamefont{Reifenberger and Tichy}(2012)}]{Reifenberger2012}
\bibinfo{author}{\bibfnamefont{G.}~\bibnamefont{Reifenberger}}
  \bibnamefont{and} \bibinfo{author}{\bibfnamefont{W.}~\bibnamefont{Tichy}},
  \bibinfo{journal}{Phys. Rev. D} \textbf{\bibinfo{volume}{86}},
  \bibinfo{pages}{064003} (\bibinfo{year}{2012}),
  \urlprefix\url{http://link.aps.org/doi/10.1103/PhysRevD.86.064003}.

\bibitem[{\citenamefont{Pfeiffer et~al.}(2003)\citenamefont{Pfeiffer, Kidder,
  Scheel, and Teukolsky}}]{Pfeiffer2003}
\bibinfo{author}{\bibfnamefont{H.~P.} \bibnamefont{Pfeiffer}},
  \bibinfo{author}{\bibfnamefont{L.~E.} \bibnamefont{Kidder}},
  \bibinfo{author}{\bibfnamefont{M.~A.} \bibnamefont{Scheel}},
  \bibnamefont{and} \bibinfo{author}{\bibfnamefont{S.~A.}
  \bibnamefont{Teukolsky}}, \bibinfo{journal}{Comput.\ Phys.\ Commun.}
  \textbf{\bibinfo{volume}{152}}, \bibinfo{pages}{253} (\bibinfo{year}{2003}).

\bibitem[{\citenamefont{Szilagyi et~al.}(2009)\citenamefont{Szilagyi, Lindblom,
  and Scheel}}]{Szilagyi:2009qz}
\bibinfo{author}{\bibfnamefont{B.}~\bibnamefont{Szilagyi}},
  \bibinfo{author}{\bibfnamefont{L.}~\bibnamefont{Lindblom}}, \bibnamefont{and}
  \bibinfo{author}{\bibfnamefont{M.~A.} \bibnamefont{Scheel}},
  \bibinfo{journal}{Phys.\ Rev.\ D} \textbf{\bibinfo{volume}{80}},
  \bibinfo{pages}{124010} (\bibinfo{year}{2009}), \eprint{0909.3557}.

\bibitem[{\citenamefont{York}(1999)}]{York1999}
\bibinfo{author}{\bibfnamefont{J.~W.} \bibnamefont{York}},
  \bibinfo{journal}{Phys.\ Rev.\ Lett.} \textbf{\bibinfo{volume}{82}},
  \bibinfo{pages}{1350} (\bibinfo{year}{1999}).

\bibitem[{\citenamefont{Pfeiffer and York}(2003)}]{Pfeiffer2003b}
\bibinfo{author}{\bibfnamefont{H.~P.} \bibnamefont{Pfeiffer}} \bibnamefont{and}
  \bibinfo{author}{\bibfnamefont{J.~W.} \bibnamefont{York}},
  \bibinfo{journal}{Phys.\ Rev.\ D} \textbf{\bibinfo{volume}{67}},
  \bibinfo{pages}{044022} (\bibinfo{year}{2003}).

\bibitem[{\citenamefont{Arnowitt et~al.}(1962)\citenamefont{Arnowitt, Deser,
  and Misner}}]{ADM}
\bibinfo{author}{\bibfnamefont{R.}~\bibnamefont{Arnowitt}},
  \bibinfo{author}{\bibfnamefont{S.}~\bibnamefont{Deser}}, \bibnamefont{and}
  \bibinfo{author}{\bibfnamefont{C.~W.} \bibnamefont{Misner}}, in
  \emph{\bibinfo{booktitle}{Gravitation: An Introduction to Current Research}},
  edited by \bibinfo{editor}{\bibfnamefont{L.}~\bibnamefont{Witten}}
  (\bibinfo{publisher}{Wiley}, \bibinfo{address}{New York},
  \bibinfo{year}{1962}).

\bibitem[{\citenamefont{{York, Jr.}}(1979)}]{york79}
\bibinfo{author}{\bibfnamefont{J.~W.} \bibnamefont{{York, Jr.}}}, in
  \emph{\bibinfo{booktitle}{Sources of Gravitational Radiation}}, edited by
  \bibinfo{editor}{\bibfnamefont{L.~L.} \bibnamefont{Smarr}}
  (\bibinfo{publisher}{Cambridge University Press},
  \bibinfo{address}{Cambridge, England}, \bibinfo{year}{1979}), pp.
  \bibinfo{pages}{83--126}.

\bibitem[{\citenamefont{Bowen and {York, Jr.}}(1980)}]{Bowen-York:1980}
\bibinfo{author}{\bibfnamefont{J.~M.} \bibnamefont{Bowen}} \bibnamefont{and}
  \bibinfo{author}{\bibfnamefont{J.~W.} \bibnamefont{{York, Jr.}}},
  \bibinfo{journal}{Phys.\ Rev.\ D} \textbf{\bibinfo{volume}{21}},
  \bibinfo{pages}{2047} (\bibinfo{year}{1980}).

\bibitem[{\citenamefont{{M. Scheel, M. Boyle, T. Chu, L. Kidder, K. Matthews
  and H. Pfeiffer}}(2009)}]{Scheel2009}
\bibinfo{author}{\bibnamefont{{M. Scheel, M. Boyle, T. Chu, L. Kidder, K.
  Matthews and H. Pfeiffer}}}, \bibinfo{journal}{Phys.\ Rev.\ D}
  \textbf{\bibinfo{volume}{79}}, \bibinfo{pages}{024003}
  (\bibinfo{year}{2009}), \eprint{arXiv:gr-qc/0810.1767}.

\bibitem[{\citenamefont{Lichnerowicz}(1944)}]{Lichnerowicz44}
\bibinfo{author}{\bibfnamefont{A.}~\bibnamefont{Lichnerowicz}},
  \bibinfo{journal}{J. Math Pures et Appl.} \textbf{\bibinfo{volume}{23}},
  \bibinfo{pages}{37} (\bibinfo{year}{1944}).

\bibitem[{\citenamefont{Smarr and York}(1978)}]{Smarr78b}
\bibinfo{author}{\bibfnamefont{L.}~\bibnamefont{Smarr}} \bibnamefont{and}
  \bibinfo{author}{\bibfnamefont{J.~W.} \bibnamefont{York}},
  \bibinfo{journal}{Phys.\ Rev.\ D} \textbf{\bibinfo{volume}{17}},
  \bibinfo{pages}{2529} (\bibinfo{year}{1978}).

\bibitem[{\citenamefont{Cook and Pfeiffer}(2004)}]{Cook2004}
\bibinfo{author}{\bibfnamefont{G.~B.} \bibnamefont{Cook}} \bibnamefont{and}
  \bibinfo{author}{\bibfnamefont{H.~P.} \bibnamefont{Pfeiffer}},
  \bibinfo{journal}{Phys.\ Rev.\ D} \textbf{\bibinfo{volume}{70}},
  \bibinfo{pages}{104016} (\bibinfo{year}{2004}).

\bibitem[{\citenamefont{{Caudill} et~al.}(2006)\citenamefont{{Caudill}, {Cook},
  {Grigsby}, and {Pfeiffer}}}]{Caudill-etal:2006}
\bibinfo{author}{\bibfnamefont{M.}~\bibnamefont{{Caudill}}},
  \bibinfo{author}{\bibfnamefont{G.~B.} \bibnamefont{{Cook}}},
  \bibinfo{author}{\bibfnamefont{J.~D.} \bibnamefont{{Grigsby}}},
  \bibnamefont{and} \bibinfo{author}{\bibfnamefont{H.~P.}
  \bibnamefont{{Pfeiffer}}}, \bibinfo{journal}{Phys.\ Rev.\ D}
  \textbf{\bibinfo{volume}{74}}, \bibinfo{pages}{064011}
  (\bibinfo{year}{2006}).

\bibitem[{\citenamefont{Pfeiffer et~al.}(2007)\citenamefont{Pfeiffer, Brown,
  Kidder, Lindblom, Lovelace, and Scheel}}]{Pfeiffer-Brown-etal:2007}
\bibinfo{author}{\bibfnamefont{H.~P.} \bibnamefont{Pfeiffer}},
  \bibinfo{author}{\bibfnamefont{D.~A.} \bibnamefont{Brown}},
  \bibinfo{author}{\bibfnamefont{L.~E.} \bibnamefont{Kidder}},
  \bibinfo{author}{\bibfnamefont{L.}~\bibnamefont{Lindblom}},
  \bibinfo{author}{\bibfnamefont{G.}~\bibnamefont{Lovelace}}, \bibnamefont{and}
  \bibinfo{author}{\bibfnamefont{M.~A.} \bibnamefont{Scheel}},
  \bibinfo{journal}{Class.\ Quantum Grav.} \textbf{\bibinfo{volume}{24}},
  \bibinfo{pages}{S59} (\bibinfo{year}{2007}).

\bibitem[{\citenamefont{Lovelace et~al.}(2012)\citenamefont{Lovelace, Boyle,
  Scheel, and Szil\'{a}gyi}}]{Lovelace:2011nu}
\bibinfo{author}{\bibfnamefont{G.}~\bibnamefont{Lovelace}},
  \bibinfo{author}{\bibfnamefont{M.}~\bibnamefont{Boyle}},
  \bibinfo{author}{\bibfnamefont{M.~A.} \bibnamefont{Scheel}},
  \bibnamefont{and}
  \bibinfo{author}{\bibfnamefont{B.}~\bibnamefont{Szil\'{a}gyi}},
  \bibinfo{journal}{Class.Quant.Grav.} \textbf{\bibinfo{volume}{29}},
  \bibinfo{pages}{045003} (\bibinfo{year}{2012}), \eprint{arXiv:1110.2229}.

\bibitem[{\citenamefont{Cook and Scheel}(1997)}]{cook_scheel97}
\bibinfo{author}{\bibfnamefont{G.~B.} \bibnamefont{Cook}} \bibnamefont{and}
  \bibinfo{author}{\bibfnamefont{M.~A.} \bibnamefont{Scheel}},
  \bibinfo{journal}{Phys.\ Rev.\ D} \textbf{\bibinfo{volume}{56}},
  \bibinfo{pages}{4775} (\bibinfo{year}{1997}).

\bibitem[{\citenamefont{Blanchet et~al.}(1998)\citenamefont{Blanchet, Faye, and
  Ponsot}}]{Blanchet-Faye-Ponsot:1998}
\bibinfo{author}{\bibfnamefont{L.}~\bibnamefont{Blanchet}},
  \bibinfo{author}{\bibfnamefont{G.}~\bibnamefont{Faye}}, \bibnamefont{and}
  \bibinfo{author}{\bibfnamefont{B.}~\bibnamefont{Ponsot}},
  \bibinfo{journal}{Phys.\ Rev.\ D} \textbf{\bibinfo{volume}{58}},
  \bibinfo{pages}{124002} (\bibinfo{year}{1998}).

\bibitem[{\citenamefont{Booth}(2005)}]{Booth2005}
\bibinfo{author}{\bibfnamefont{I.}~\bibnamefont{Booth}}, \bibinfo{journal}{Can.
  J. Phys.} \textbf{\bibinfo{volume}{83}}, \bibinfo{pages}{1073}
  (\bibinfo{year}{2005}), \eprint{gr-qc/0508107}.

\bibitem[{\citenamefont{Johnson-McDaniel}(2012)}]{Nathan:Email:July27-2012}
\bibinfo{author}{\bibfnamefont{N.~K.} \bibnamefont{Johnson-McDaniel}},
  \bibinfo{howpublished}{private communication} (\bibinfo{year}{2012}).

\bibitem[{\citenamefont{Scheel et~al.}(2006)\citenamefont{Scheel, Pfeiffer,
  Lindblom, Kidder, Rinne, and Teukolsky}}]{Scheel2006}
\bibinfo{author}{\bibfnamefont{M.~A.} \bibnamefont{Scheel}},
  \bibinfo{author}{\bibfnamefont{H.~P.} \bibnamefont{Pfeiffer}},
  \bibinfo{author}{\bibfnamefont{L.}~\bibnamefont{Lindblom}},
  \bibinfo{author}{\bibfnamefont{L.~E.} \bibnamefont{Kidder}},
  \bibinfo{author}{\bibfnamefont{O.}~\bibnamefont{Rinne}}, \bibnamefont{and}
  \bibinfo{author}{\bibfnamefont{S.~A.} \bibnamefont{Teukolsky}},
  \bibinfo{journal}{Phys.\ Rev.\ D} \textbf{\bibinfo{volume}{74}},
  \bibinfo{pages}{104006} (\bibinfo{year}{2006}).

\bibitem[{SpE()}]{SpECwebsite}
\bibinfo{howpublished}{\url{http://www.black-holes.org/SpEC.html}}.

\bibitem[{\citenamefont{Lindblom et~al.}(2006)\citenamefont{Lindblom, Scheel,
  Kidder, Owen, and Rinne}}]{Lindblom2006}
\bibinfo{author}{\bibfnamefont{L.}~\bibnamefont{Lindblom}},
  \bibinfo{author}{\bibfnamefont{M.~A.} \bibnamefont{Scheel}},
  \bibinfo{author}{\bibfnamefont{L.~E.} \bibnamefont{Kidder}},
  \bibinfo{author}{\bibfnamefont{R.}~\bibnamefont{Owen}}, \bibnamefont{and}
  \bibinfo{author}{\bibfnamefont{O.}~\bibnamefont{Rinne}},
  \bibinfo{journal}{Class.\ Quantum Grav.} \textbf{\bibinfo{volume}{23}},
  \bibinfo{pages}{S447} (\bibinfo{year}{2006}).

\bibitem[{\citenamefont{Friedrich}(1985)}]{Friedrich1985}
\bibinfo{author}{\bibfnamefont{H.}~\bibnamefont{Friedrich}},
  \bibinfo{journal}{Commun.\ Math.\ Phys.} \textbf{\bibinfo{volume}{100}},
  \bibinfo{pages}{525} (\bibinfo{year}{1985}).

\bibitem[{\citenamefont{Garfinkle}(2002)}]{Garfinkle2002}
\bibinfo{author}{\bibfnamefont{D.}~\bibnamefont{Garfinkle}},
  \bibinfo{journal}{Phys.\ Rev.\ D} \textbf{\bibinfo{volume}{65}},
  \bibinfo{pages}{044029} (\bibinfo{year}{2002}).

\bibitem[{\citenamefont{Pretorius}(2005)}]{Pretorius2005c}
\bibinfo{author}{\bibfnamefont{F.}~\bibnamefont{Pretorius}},
  \bibinfo{journal}{Class.\ Quantum Grav.} \textbf{\bibinfo{volume}{22}},
  \bibinfo{pages}{425} (\bibinfo{year}{2005}).

\bibitem[{\citenamefont{Rinne}(2006)}]{Rinne2006}
\bibinfo{author}{\bibfnamefont{O.}~\bibnamefont{Rinne}},
  \bibinfo{journal}{Class.\ Quantum Grav.} \textbf{\bibinfo{volume}{23}},
  \bibinfo{pages}{6275} (\bibinfo{year}{2006}).

\bibitem[{\citenamefont{Rinne et~al.}(2007)\citenamefont{Rinne, Lindblom, and
  Scheel}}]{Rinne2007}
\bibinfo{author}{\bibfnamefont{O.}~\bibnamefont{Rinne}},
  \bibinfo{author}{\bibfnamefont{L.}~\bibnamefont{Lindblom}}, \bibnamefont{and}
  \bibinfo{author}{\bibfnamefont{M.~A.} \bibnamefont{Scheel}},
  \bibinfo{journal}{Class.\ Quantum Grav.} \textbf{\bibinfo{volume}{24}},
  \bibinfo{pages}{4053} (\bibinfo{year}{2007}).

\bibitem[{\citenamefont{Stewart}(1998)}]{Stewart1998}
\bibinfo{author}{\bibfnamefont{J.~M.} \bibnamefont{Stewart}},
  \bibinfo{journal}{Class.\ Quantum Grav.} \textbf{\bibinfo{volume}{15}},
  \bibinfo{pages}{2865} (\bibinfo{year}{1998}).

\bibitem[{\citenamefont{Friedrich and Nagy}(1999)}]{FriedrichNagy1999}
\bibinfo{author}{\bibfnamefont{H.}~\bibnamefont{Friedrich}} \bibnamefont{and}
  \bibinfo{author}{\bibfnamefont{G.}~\bibnamefont{Nagy}},
  \bibinfo{journal}{Commun.\ Math.\ Phys.} \textbf{\bibinfo{volume}{201}},
  \bibinfo{pages}{619} (\bibinfo{year}{1999}).

\bibitem[{\citenamefont{Bardeen and Buchman}(2002)}]{Bardeen2002}
\bibinfo{author}{\bibfnamefont{J.~M.} \bibnamefont{Bardeen}} \bibnamefont{and}
  \bibinfo{author}{\bibfnamefont{L.~T.} \bibnamefont{Buchman}},
  \bibinfo{journal}{Phys.\ Rev.\ D} \textbf{\bibinfo{volume}{65}},
  \bibinfo{pages}{064037} (\bibinfo{year}{2002}).

\bibitem[{\citenamefont{Szil\'agyi et~al.}(2002)\citenamefont{Szil\'agyi,
  Schmidt, and Winicour}}]{Szilagyi2002}
\bibinfo{author}{\bibfnamefont{B.}~\bibnamefont{Szil\'agyi}},
  \bibinfo{author}{\bibfnamefont{B.}~\bibnamefont{Schmidt}}, \bibnamefont{and}
  \bibinfo{author}{\bibfnamefont{J.}~\bibnamefont{Winicour}},
  \bibinfo{journal}{Phys.\ Rev.\ D} \textbf{\bibinfo{volume}{65}},
  \bibinfo{pages}{064015} (\bibinfo{year}{2002}).

\bibitem[{\citenamefont{{Calabrese} et~al.}(2003)\citenamefont{{Calabrese},
  {Pullin}, {Reula}, {Sarbach}, and {Tiglio}}}]{Calabrese2003}
\bibinfo{author}{\bibfnamefont{G.}~\bibnamefont{{Calabrese}}},
  \bibinfo{author}{\bibfnamefont{J.}~\bibnamefont{{Pullin}}},
  \bibinfo{author}{\bibfnamefont{O.}~\bibnamefont{{Reula}}},
  \bibinfo{author}{\bibfnamefont{O.}~\bibnamefont{{Sarbach}}},
  \bibnamefont{and} \bibinfo{author}{\bibfnamefont{M.}~\bibnamefont{{Tiglio}}},
  \bibinfo{journal}{Commun.\ Math.\ Phys.} \textbf{\bibinfo{volume}{240}},
  \bibinfo{pages}{377} (\bibinfo{year}{2003}), \eprint{gr-qc/0209017}.

\bibitem[{\citenamefont{Szil\'agyi and Winicour}(2003)}]{Szilagyi2003}
\bibinfo{author}{\bibfnamefont{B.}~\bibnamefont{Szil\'agyi}} \bibnamefont{and}
  \bibinfo{author}{\bibfnamefont{J.}~\bibnamefont{Winicour}},
  \bibinfo{journal}{Phys.\ Rev.\ D} \textbf{\bibinfo{volume}{68}},
  \bibinfo{pages}{041501(R)} (\bibinfo{year}{2003}).

\bibitem[{\citenamefont{Kidder et~al.}(2005)\citenamefont{Kidder, Lindblom,
  Scheel, Buchman, and Pfeiffer}}]{Kidder2005}
\bibinfo{author}{\bibfnamefont{L.~E.} \bibnamefont{Kidder}},
  \bibinfo{author}{\bibfnamefont{L.}~\bibnamefont{Lindblom}},
  \bibinfo{author}{\bibfnamefont{M.~A.} \bibnamefont{Scheel}},
  \bibinfo{author}{\bibfnamefont{L.~T.} \bibnamefont{Buchman}},
  \bibnamefont{and} \bibinfo{author}{\bibfnamefont{H.~P.}
  \bibnamefont{Pfeiffer}}, \bibinfo{journal}{Phys.\ Rev.\ D}
  \textbf{\bibinfo{volume}{71}}, \bibinfo{pages}{064020}
  (\bibinfo{year}{2005}).

\bibitem[{\citenamefont{Buchman and Sarbach}(2006)}]{Buchman2006}
\bibinfo{author}{\bibfnamefont{L.~T.} \bibnamefont{Buchman}} \bibnamefont{and}
  \bibinfo{author}{\bibfnamefont{O.~C.~A.} \bibnamefont{Sarbach}},
  \bibinfo{journal}{Class.\ Quantum Grav.} \textbf{\bibinfo{volume}{23}},
  \bibinfo{pages}{6709} (\bibinfo{year}{2006}).

\bibitem[{\citenamefont{Buchman and Sarbach}(2007)}]{Buchman2007}
\bibinfo{author}{\bibfnamefont{L.~T.} \bibnamefont{Buchman}} \bibnamefont{and}
  \bibinfo{author}{\bibfnamefont{O.~C.~A.} \bibnamefont{Sarbach}},
  \bibinfo{journal}{Class.\ Quantum Grav.} \textbf{\bibinfo{volume}{24}},
  \bibinfo{pages}{S307} (\bibinfo{year}{2007}).

\bibitem[{\citenamefont{Gottlieb and Hesthaven}(2001)}]{Gottlieb2001}
\bibinfo{author}{\bibfnamefont{D.}~\bibnamefont{Gottlieb}} \bibnamefont{and}
  \bibinfo{author}{\bibfnamefont{J.~S.} \bibnamefont{Hesthaven}},
  \bibinfo{journal}{J. Comput. Appl. Math.} \textbf{\bibinfo{volume}{128}},
  \bibinfo{pages}{83} (\bibinfo{year}{2001}), ISSN \bibinfo{issn}{0377-0427}.

\bibitem[{\citenamefont{Hesthaven}(2000)}]{Hesthaven2000}
\bibinfo{author}{\bibfnamefont{J.~S.} \bibnamefont{Hesthaven}},
  \bibinfo{journal}{Appl. Num. Math.} \textbf{\bibinfo{volume}{33}},
  \bibinfo{pages}{23} (\bibinfo{year}{2000}).

\bibitem[{\citenamefont{Lindblom et~al.}(2008)\citenamefont{Lindblom, Matthews,
  Rinne, and Scheel}}]{Lindblom2007}
\bibinfo{author}{\bibfnamefont{L.}~\bibnamefont{Lindblom}},
  \bibinfo{author}{\bibfnamefont{K.~D.} \bibnamefont{Matthews}},
  \bibinfo{author}{\bibfnamefont{O.}~\bibnamefont{Rinne}}, \bibnamefont{and}
  \bibinfo{author}{\bibfnamefont{M.~A.} \bibnamefont{Scheel}},
  \bibinfo{journal}{Phys.\ Rev.\ D} \textbf{\bibinfo{volume}{77}},
  \bibinfo{pages}{084001} (\bibinfo{year}{2008}).

\bibitem[{\citenamefont{Hawking}(1968)}]{Hawking1968}
\bibinfo{author}{\bibfnamefont{S.~W.} \bibnamefont{Hawking}},
  \bibinfo{journal}{J.\ Math.\ Phys.} \textbf{\bibinfo{volume}{9}},
  \bibinfo{pages}{598} (\bibinfo{year}{1968}).

\bibitem[{\citenamefont{Taylor and Poisson}(2008)}]{Poisson2008}
\bibinfo{author}{\bibfnamefont{S.}~\bibnamefont{Taylor}} \bibnamefont{and}
  \bibinfo{author}{\bibfnamefont{E.}~\bibnamefont{Poisson}},
  \bibinfo{journal}{Phys.\ Rev.\ D} \textbf{\bibinfo{volume}{78}},
  \bibinfo{pages}{084016} (\bibinfo{year}{2008}).

\bibitem[{\citenamefont{Yunes}(2007)}]{Yunes2007}
\bibinfo{author}{\bibfnamefont{N.}~\bibnamefont{Yunes}},
  \bibinfo{journal}{Class.\ Quantum Grav.} \textbf{\bibinfo{volume}{24}},
  \bibinfo{pages}{4313} (\bibinfo{year}{2007}), \eprint{0611128}.

\bibitem[{\citenamefont{Throwe et~al.}()}]{ThroweInPrep}
\bibinfo{author}{\bibfnamefont{W.}~\bibnamefont{Throwe}} \bibnamefont{et~al.},
  \bibinfo{note}{in preparation}.

\bibitem[{\citenamefont{MacDonald et~al.}(2013)\citenamefont{MacDonald,
  Mrou\'e, Pfeiffer, Boyle, Kidder, Scheel, Szil{\' a}gyi, and
  Taylor}}]{MacDonald:2012mp}
\bibinfo{author}{\bibfnamefont{I.}~\bibnamefont{MacDonald}},
  \bibinfo{author}{\bibfnamefont{A.~H.} \bibnamefont{Mrou\'e}},
  \bibinfo{author}{\bibfnamefont{H.~P.} \bibnamefont{Pfeiffer}},
  \bibinfo{author}{\bibfnamefont{M.}~\bibnamefont{Boyle}},
  \bibinfo{author}{\bibfnamefont{L.~E.} \bibnamefont{Kidder}},
  \bibinfo{author}{\bibfnamefont{M.~A.} \bibnamefont{Scheel}},
  \bibinfo{author}{\bibfnamefont{B.}~\bibnamefont{Szil{\' a}gyi}},
  \bibnamefont{and} \bibinfo{author}{\bibfnamefont{N.~W.}
  \bibnamefont{Taylor}}, \bibinfo{journal}{Phys.\ Rev.\ D}
  \textbf{\bibinfo{volume}{87}}, \bibinfo{pages}{024009}
  (\bibinfo{year}{2013}), \eprint{1210.3007}.

\bibitem[{\citenamefont{Winicour}(2009)}]{Winicour2009}
\bibinfo{author}{\bibfnamefont{J.}~\bibnamefont{Winicour}},
  \bibinfo{journal}{Living Rev.~Rel.} \textbf{\bibinfo{volume}{12}}
  (\bibinfo{year}{2009}),
  \urlprefix\url{http://www.livingreviews.org/lrr-2009-3}.

\bibitem[{\citenamefont{Taylor et~al.}(2013)\citenamefont{Taylor, Boyle,
  Reisswig, Scheel, Chu et~al.}}]{Taylor:2013zia}
\bibinfo{author}{\bibfnamefont{N.~W.} \bibnamefont{Taylor}},
  \bibinfo{author}{\bibfnamefont{M.}~\bibnamefont{Boyle}},
  \bibinfo{author}{\bibfnamefont{C.}~\bibnamefont{Reisswig}},
  \bibinfo{author}{\bibfnamefont{M.~A.} \bibnamefont{Scheel}},
  \bibinfo{author}{\bibfnamefont{T.}~\bibnamefont{Chu}}, \bibnamefont{et~al.}
  (\bibinfo{year}{2013}), \bibinfo{note}{arXiv:1309.3605}, \eprint{1309.3605}.

\bibitem[{\citenamefont{Thorne and Hartle}(1985)}]{Thorne1985}
\bibinfo{author}{\bibfnamefont{K.~S.} \bibnamefont{Thorne}} \bibnamefont{and}
  \bibinfo{author}{\bibfnamefont{J.~B.} \bibnamefont{Hartle}},
  \bibinfo{journal}{Phys.\ Rev.\ D} \textbf{\bibinfo{volume}{31}},
  \bibinfo{pages}{1815} (\bibinfo{year}{1985}).

\end{thebibliography}
\end{document}